\documentclass{article}

\usepackage{graphics}
\usepackage{graphicx}
\usepackage{wrapfig}
\usepackage{hyperref}
\usepackage{subcaption}
\usepackage{amsfonts}
\usepackage{algorithm,algorithmicx}
\usepackage{algpseudocode}
\usepackage{microtype}      
\usepackage{pdfpages}
\usepackage{multicol}
\usepackage{amsmath}
\usepackage{amssymb}
\usepackage{float}
\usepackage{amsthm}
\usepackage[toc,page]{appendix}

\usepackage[accepted]{icml2021}

\icmltitlerunning{TradeR: Practical Deep Hierarchical Reinforcement Learning for Trade Execution}

\begin{document}

\twocolumn[
\icmltitle{TradeR: Practical Deep Hierarchical Reinforcement Learning for Trade Execution}



\icmlsetsymbol{equal}{*}

\begin{icmlauthorlist}
\icmlauthor{Karush Suri}{to,goo}
\icmlauthor{Xiao Qi Shi}{ed}
\icmlauthor{Konstantinos Plataniotis}{to}
\icmlauthor{Yuri Lawryshyn}{goo}

\end{icmlauthorlist}

\icmlaffiliation{to}{Department of Electrical \& Computer Engineering, University of Toronto, Canada.}
\icmlaffiliation{goo}{Center for Management of Technology \& Entrepreneurship (CMTE)}
\icmlaffiliation{ed}{RBC Capital Markets}

\icmlcorrespondingauthor{Karush Suri}{karush.suri@mail.utoronto.ca}

\icmlkeywords{Hierarchical, Surprise Minimization, Reinforcement Learning, TradeR}

\vskip 0.3in
]



\printAffiliationsAndNotice{} 

\begin{abstract}
    Advances in Reinforcement Learning (RL) span a wide variety of applications which motivate development in this area. While application tasks serve as suitable benchmarks for real world problems, RL is seldomly used in practical scenarios consisting of abrupt dynamics. This allows one to rethink the problem setup in light of practical challenges. We present Trade Execution using Reinforcement Learning (TradeR) which aims to address two such practical challenges of \textit{catastrophy} and \textit{surprise minimization} by formulating trading as a real-world hierarchical RL problem. Through this lens, TradeR makes use of hierarchical RL to execute trade bids on high-frequency real market experiences comprising of abrupt price variations during the 2019 fiscal year COVID19 stock market crash. The framework utilizes an energy-based scheme in conjunction with surprise value function for estimating and minimizing surprise. In a large-scale study of 35 stock symbols from the S\&P500 index, TradeR demonstrates robustness to abrupt price changes and catastrophic losses while maintaining profitable outcomes. We hope that our work serves as a motivating example for application of RL to practical problems.

\end{abstract}

\section{Introduction}
RL has demonstrated profound success in applications such as video games \cite{dqn}, board games \cite{go} and robot control \cite{ppo, ddpg}. Primary reason for increasing developments in RL is the provision of deep neural networks as efficient function approximators. Combination of non-linear approximations of action distributions in the agent-environment paradigm allows for perception of large state spaces which otherwise hinder scalability of the control framework. While applications of RL span a wide variety of tasks which serve as test benchmarks for real world problems, RL is seldomly used for practical applications \cite{loon}. 

Consider the problem of autonomous driving wherein a system autonomously operates a vehicle in real-time. Formulating this as a reinforcement learning problem may be carried out by treating the system as \textit{agent} and a simulator as \textit{environment}. At each step, the system observes pixel inputs as \textit{states} from the simulator and outputs steering, acceleration and brake commands as \textit{actions}. The simulator consists of a reward function which depends on the distance covered by the vehicle and its speed under different traffic conditions. Additionally, the reward consists of a penalty if the system collides with any obstacles in the simulator, for instance another vehicle or pedestrians. The system learns to optimize the reward function by maintaining constant speed and covering more distance in different traffic conditions. Whenever the vehicle approaches an obstacle, the system acts to avoid it by utilizing the brake and directional steering commands. Such behaviors acquired by the system during training denote promise for practical adaptation. 

However, the framework cannot be adapted in real-world traffic for two main reasons. Firstly, the system would learn from trial and error leading to frequent mistakes causing damage to the vehicle or obstacles. In general, the \textit{agent} cannot carry out \textit{catastrophy minimization} quickly and efficiently. Secondly, irrespective of obstacle penalty in the reward function, the system may abruptly encounter an obstacle (a child running to cross the street) resulting in collision. Even though the \textit{agent} tries to act optimally (pushing the brakes), it fails to evade a collision as a result of the fast-paced dynamics. As a result, the \textit{agent} cannot carry out \textit{surprise minimization} betweeen consecutive states. 

While various open problems such as data efficiency \cite{offline} and estimation bias \cite{double-q-learning,ddqn} hinder an agent's performance, \textit{catastrophy} and \textit{surprise minimization} hinder agent's scalability to real world scenarios. Catastrophic and surprising states scale linearly with the state space of the environment making it increasingly challenging for the agent to quickly switch between actions. Furthermore, the occurence of such abrupt transitions is rarely encountered during the training phase resulting in a lack of informative data collection. This highlights the necessity to approach RL from a realistic viewpoint. 

In order to address the aforementioned challenges from a practical standpoint, we turn our attention to the more challenging scenario of trading. The \textit{agent} (trader) interacts in the \textit{environment} (market) to maximize its \textit{reward} (profit). We propose Trade Execution using Reinforcement Learning (TradeR), a hierarchical framework for executing trade bids on abrupt real market experiences during the COVID19 stock market crash. TradeR fulfills \textit{catastrophy minimization} by providing the agent with order and bid policies capable of estimating order quantities and executing bid actions respectively. The order policy determines the quantity for the bid. The high-level bid policy, utilizes the estimated quantity in conjunction with current states to execute actions as trade bids. TradeR further tackles \textit{surprise minimization} by generalizing an energy-based scheme \cite{emix} comprising of a surprise value function. The value function estimates surprise correponding to a given state which is inferred by the agent as intrinsic motivation. In our high-frequency and large-scale study of 35 stock symbols from the S\&P500 index for 2019 fiscal year, TradeR demonstrates improved robustness to abrupt changes with consistent profitable outcomes. 

\section{Preliminaries}
We review the RL setup wherein an agent interacts with the environment in order to transition to new states and observe rewards by following a sequence of actions. The problem is modeled as a finite-horizon Markov Decision Process (MDP) \citep{rl} defined by the tuple $(\mathcal{S},\mathcal{A},r,P,\gamma)$ where the state space is denoted by $\mathcal{S}$ and action space by $\mathcal{A}$, $r$ presents the reward observed by agent such that $r: \mathcal{S} \times \mathcal{A} \rightarrow [r_{min},r_{max}]$, $P: \mathcal{S} \times \mathcal{S} \times \mathcal{A} \rightarrow [0,\infty)$ presents the unknown transition model consisting of the transition probability to the next state $s_{t+1} \in \mathcal{S}$ given the current state $s_{t} \in \mathcal{S}$ and action $a_{t} \in \mathcal{A}$ at time step $t$ and $\gamma$ is the discount factor. We consider a policy $\pi_{\theta}({a_{t}|s_{t}})$ as a function of model parameters $\theta$. 

Standard RL defines the agent's objective to maximize the expected discounted reward $\mathbb{E}_{\pi_{\theta}}[\sum_{t=0}^{T}\gamma^{t}r(s_{t},a_{t})]$ as a function of the parameters $\theta$. 
For a fixed policy, the $Q$-value function can be computed iteratively, starting from any function $Q:\mathcal{S} \times \mathcal{A}$ and repeatedly applying a modified Bellman backup operator $\mathcal{T}^{\pi}$ given by $\mathcal{T}^{\pi}Q(s_{t},a_{t}) = r(s_{t},a_{t}) + \gamma\mathbb{E}_{s_{t+1} \sim P}[V(s_{t+1})]$ where $V(s_{t}) = \mathbb{E}_{a_{t} \sim \pi}[Q(s_{t},a_{t})]$ is the state value function. We consider a parameterized state value function $V_{\psi}(s_{t})$ and a policy $\pi_{\theta}(a_{t}|s_{t})$ which can be represented with neural networks with $\psi$ and $\theta$ being the parameters of these networks.

\section{Related Work}
\subsection{Trade Execution using RL}
 Various prior works in literature have adopted trading as a suitable task for developing practical RL algorithms. \cite{moody} presents the first RL trading framework capable of training Q-learning and recurrent RL agents based on financial metrics such as Sharpe Ratio. The work of \cite{pracddpg} provides a framework for Deterministic Policy Gradients in optimizing daily trades using additional financial metrics and cumulative returns. \cite{intraday} extends the RL trading strategy towards intraday execution comprising of asynchronous distributed updates in conjunction with backtesting methods. The resulting method is found to be sample efficient. \cite{appq} and \cite{algo} reduce the need for financial metrics during evaluation as a result of optimization of cumulative portfolio returns. The framework of portfolio optimization is further extended to cryptocurrency based on memory vectors \cite{mem} and event-based transitions \cite{event}. \cite{ddpo} generalizes these methods by retrieving the optimal strategy using deterministic portfolio optimization under varying dynamics. 
 
 While most methods focus on the application of RL to the trading problem, \cite{ensemble} improves the practicality of RL framework utilizing an ensemble of policy-based agents from the prior work of \cite{pracddpg}. Our framework for improving practical RL in trading is parallel to \cite{ensemble} and \cite{pracddpg}. 

\subsection{Hierarchical RL}
Various RL pipelines \cite{hdqn, serg1, serg2, options, temp, maxq, abbeel, feudal, esac} leverage hierarchical structure in policies allowing the agent to address different sequential problems. \cite{temp} presents the primary framework of temporal abstraction for training the policy at varying time scales consisting of sequential MDPs. \cite{hdqn} extends the temporal abstraction algorithm by incorporating intrinsic motivation by virtue of fulfilled sub-goals by the agent. Another suitable technique for learning different policy levels is the MAXQ algorithm \cite{maxq} which provisions the usage of different $Q$-values at each stage of the hierarchy. However, MAXQ demonstrates exponential computational requirements in the case of increasingly complex action spaces. A more scalable alternative are Feudal Networks \cite{feudal} which provision a manager and worker network for policy abstraction. 

Various methods leverage hierarchies in order to solve sparse rewards and long-horizon tasks. \cite{serg1} demonstrates the efficacy of learning hierarchies for locomotion tasks requiring efficient exploration. \cite{serg2} highlights the need for sample efficient hierarchies in robot control using off-policy correction. \cite{abbeel} extend hierarchical RL for learning relevant skills which support continuous adaptation. Our practical extension of hierarchical RL aligns well with these prior methods.

\section{Trade Execution using Reinforcement Learning}
In this section we introduce the TradeR framework. TradeR leverages hierarchies for long-horizon trade execution and an energy-based intrinsic motivation scheme for surprise minimization.

\subsection{Hierarchical Trade Execution}
Maximization of long-term payoffs requires an agent to predict accurately towards the future. Accurate predictions steer agent's beliefs toward an optimal policy $\pi^{*}(a_{t}|s_{t})$. However,  incorporating accurate prediction mechanisms has proven to be a non-trivial challenge in the long-horizon. While bootstrapping value estimates in TD targets extends the forward-view of the agent, this small fix in itself is insufficient as timesteps tend to infinity. Another suitable alternative is to make use of $n$-step returns which trade-off high variance at the cost of biased estimates. Tuning the discount factor $\gamma$ also does little to align agent's beliefs with counterfactual states stemming from expected value estimates.  
\begin{figure}[H]
    \centering
    \includegraphics[height=4cm,width=6cm]{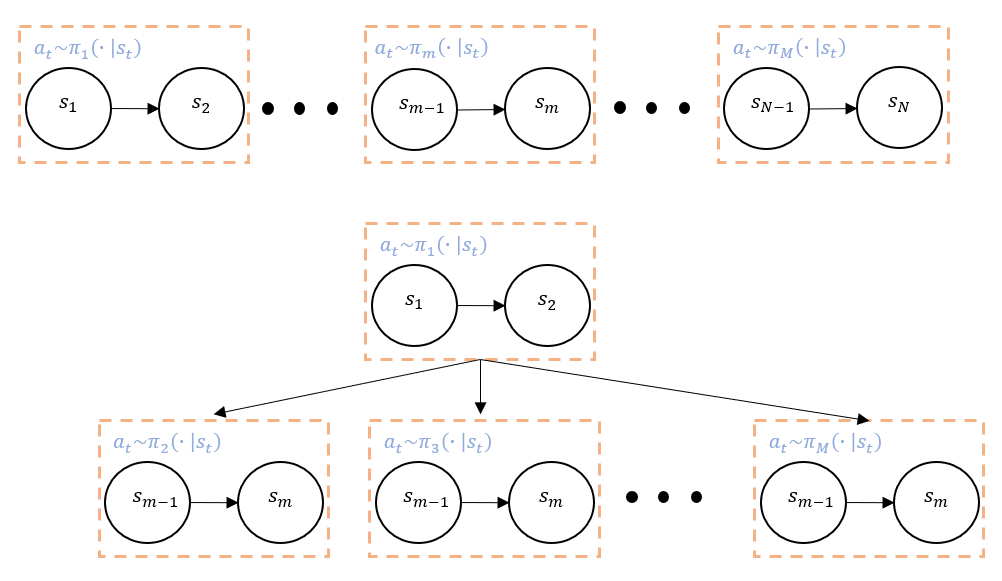}
    \caption{Sequential sub-MDPs form a Markov Chain allowing the agent to exhibit different policies based on previous ones. Simulatenous sub-MDPs form a multi-node head-to-tail graph in which policies are conditionally independent.}
    \label{fig:mdp}
\end{figure}

Hierarchical abstraction of policy into its components is a suitable mechanism which allows temporal coherency in state and value predictions. A policy $\pi(a_{t}|s_{t})$ may be decoupled into its $M$ components $(\pi_{1}(a_{t}|s_{t}), \pi_{2}(a_{t}|s_{t}),...\pi_{M}(a_{t}|s_{t}))$ where each sub-policy $\pi_{m}(a_{t}|s_{t})$ is responsible for solving a sub-task in the form of an MDP. Consequently, the joint MDP is decoupled into $M$ sub-MDPs in which the agent can interact sequentially or simulatenously. In the case of sequential interaction, the sub-MDPs together form a Markov Chain representing the joint MDP. In the case of simulatenous interactions, the joint MDP constitutes a multi-node head-to-tail graph wherein the sub-MDPs are temporally independent. This insight is presented in \autoref{fig:mdp}.   

The TradeR framework utilizes sequential interactions of policies in its sub-MDPs. The joint trading problem is abstracted into two sub-problems of (1) order determination and (2) bid execution. The agent with parameters $\theta$, jointly optimizes the objective $\mathbb{E}_{\pi}[\sum_{t=0}^{\infty}\gamma^{t}r(s_{t},a_{t})]$ since both tasks share a common reward function. Each task is executed  using a nonlinear function approximator (deep neural networks) denoting the sub-policy $\pi_{ord}(\cdot|s_{t})$ for order network and $\pi_{bid}(a_{t}|s_{t})$ for bid network respectively. We direct the reader to \autoref{appendix:implem} for details on algorithm. 
\begin{figure*}[ht]
  \centering
    \includegraphics[height=5cm,width=11cm]{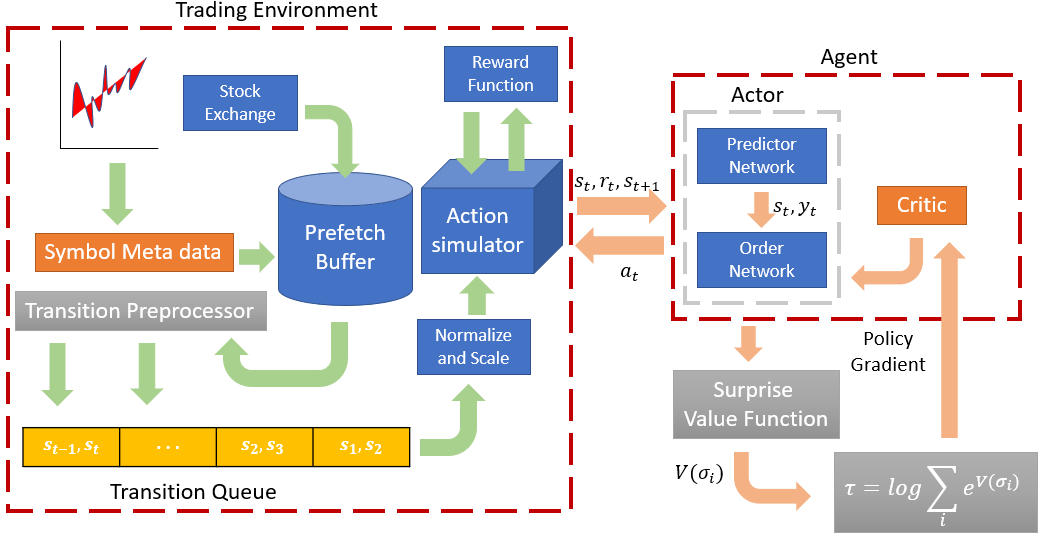}
    \caption{The data-driven environment comprises of market price transitions obtained from stock exchange. The agent, consisting of order and bid networks, minimizes catastrophy using surprise value function as part of the energy-based scheme.}
     \label{fig:TradeR}
\end{figure*}

\textbf{Order Determination:} At a given time-step $t$, the order policy observes state $s_{t}$ from the trading environment. Upon observing state $s_{t}$, the order network deterministically follows the sub-policy $\pi_{ord}(y_{t}|s_{t})$ for estimating the order quantity $y_{t}$ to be bought, sold or held by the agent. The quantity $y_{t}$ denotes the amount agent is willing to expense in the case of a loss or gain in the case of a profit.

\textbf{Bid Execution:}  The high-level bid policy observes $s_{t}$ and the quantity estimated by order policy $y_{t}$, in order to yield an action $a_{t} \sim \pi_{bid}(a_{t}|s_{t})$ denoting the bid to be executed (buy, sell or hold the quantity). Action $a_{t}$ is executed in the environment by the dual-policy framework which results in a transition to the next state $s_{t+1}$ by means of the transition probability $p(s_{t+1}|s_{t},a_{t}) \in P$. 

Order and bid policies can be updated jointly or asynchronously using either the on-policy or off-policy setting. In the on-policy case, order determination can be carried out on price samples being observed by the current policy. Bids are then executed on price samples and the corresponding quantities being estimated. In the off-policy setting, a replay buffer may be used to store experiences of quantity estimation and bid executions which are later used to update the policy. Compatibility of the framework with both on-policy and off-policy settings highlights that TradeR can be combined with any RL agent. Another notable observation is that while the action space for order determination is continuous, the bid execution task consists of discrete actions. Such a framework consisting of both continuous and discrete action spaces is representative of the practical nature of trading wherein the order quantity is prefixed in a given range and human experts make the bid once a suitable price quote is encountered.  

\subsection{Energy-based Intrinsic Motivation}
Despite the provision of hierarchical order and bid policies, the agent's policy may cripple to sub-optimal performance as a result of abrupt changes in prices. This is in agreement with modern RL applications of robot control \cite{serg1, serg2} wherein a sudden change in the position and angle of robot's placement hurts the robustness of the policy and steers it off the path towards convergence. One can leverage contraction theory \cite{banach, contraction} to highlight mathematical properties which would aid in extracting a surprise-robust policy. Such a mapping guarantees convergence in asymptotically-stable nonlinear systems \cite{contraction} and provides a dynamical framework for assessing stability of control in RL \cite{sql}. A contraction mapping between any two functions $f_{1}(w)$ and $f_{2}(w)$ implies that the norm distance between $f_{1}(w)$ and $f_{2}(w)$ decays at a constant (in some cases geometric \cite{banach}) rate. Given a contraction operator $\tau$ when iteratively applied on $f_{1}(w)$ and $f_{2}(w)$, the mapping $\tau f_{1}(w) - \tau f_{2}(w)$ is a contraction if the inequality in \autoref{eq:contraction} is satisfied.
\begin{gather}
  ||\tau f_{1}(w) - \tau f_{2}(w)|| \leq \eta ||f_{1}(w) - f_{2}(w)||,\; \forall w \; \text{s.t.}\; \eta < 1 \label{eq:contraction}
\end{gather} 
\autoref{eq:contraction} is a generalization of the fixed-point theory in Banach metric spaces \cite{contraction} which provides suitable conditions for assessing stability of nonlinear systems. The key component of evaluating a nonlinear system is motivated by its convergence towards a fixed point in the Banach metric space. Convergence towards a fixed point indicates stability of the overall mapping. 

We borrow from this insight in order to form a contraction on standard deviations in state transitions as a continuum in a nonlinear space. Upon realizing input samples as standard deviations of state transitions in this continuous nonlinear space, a simple yet elegant formulation of a contraction mapping can be achieved. To utilize a contraction mapping on $\sigma_{i}$ with $i$ being the state dimension, we seek a contraction operator $\mathcal{T}$ which is tractable. 

A suitable choice is the alternate Boltzmann (mellowmax) operator introduced in \cite{mellowmax} and presented in \autoref{eq:mellow}.
\begin{gather}
    \mathcal{T}f(w)=\log \sum_{w} \exp{f(w)} \label{eq:mellow}
\end{gather}
The Mellowmax operator can be interpreted as an energy-based function. Retaining properties of the Boltzmann distribution, \autoref{eq:mellow} is an asymptotically stable formulation of the Gibbs distribution. Mellowmax has been suitably adopted in control and learning settings \cite{sql,emix} wherein the probability distribution forms a continuum over the input space. Additionally, the exponent in $\mathcal{T}$ is tractable as it is followed by the $\log$ which prevents the arguments from exploding.
\begin{gather}
  \hat{r}(s_{t},a_{t},\sigma_{i}) = r(s_{t},a_{t}) + \log \sum_{i} \exp{(V_{surp}(\sigma_{i}))} \label{eq:energy}
\end{gather}
\autoref{eq:energy} presents the modified reward signal utilized for surprise minimization as an intrinsic motivation objective. We first encode the stanard deviations $\sigma_{i}$ corresponding to each dimension of state space $i$ using a surprise value function $V_{surp}(\sigma_{i})$. The surprise value function $V_{surp}(\sigma_{i})$ quantifies the degree of surprise encountered by the agent in past states of the batch based on transition dynamics of the MDP. Note that $V_{surp}(\sigma_{i})$ does not build a model of the world and it does not have access to state space dynamics. The surprise value function serves as \textit{surprise critic} guiding the agent away from surprising states. \autoref{fig:TradeR} presents the complete TradeR framework. Application of surprise value function is followed by the mellowmax operator which estimates nonlinear surprising configurations corresponding to actions of the agent. This energy-based estimate of surprise is coupled with reward $r(s_{t},a_{t})$ as intrinsic motivation in order to yield the surprise-agnostic reward signal $\hat{r}(s_{t},a_{t})$. 

\subsection{The TradeR Algorithm}
\begin{algorithm}[H]
    \caption{TradeR}
    \label{alg:algorithm1}
    \begin{algorithmic}[1]
    
      \State Initialize policy $\pi_{\theta} = (\pi_{ord}, \pi_{bid})$ using $\theta$ and surprise value function $V_{surp}$ using $\phi$.
      \State Initialize learning rate $\alpha$, temperature $\beta$ and replay buffer $\mathcal{R}$.
      \State Initialize environment with initial state $s_{1}$
    
      \For{environment step}
          \State Sample quantity $y_{t} \sim \pi_{ord}(\cdot|s_{t})$
          \State Sample bid $a_{t} \sim \pi_{bid}(\cdot|s_{t},y_{t})$
          \State Execute $a_{t}$ and observe $s_{t+1}$ and $r_{t}$
          \State Update buffer $\mathcal{R} = \mathcal{R} \cup (s_{t},a_{t},r_{t},s_{t+1})$
      \EndFor
      \For{random batch}
          \State Compute deviations $\sigma_{i}$ in states
          \State Set reward $\hat{r}(s_{t},a_{t},\sigma_{t}) = r(s_{t},a_{t}) + \beta \log \sum_{i}\exp (V_{surp}(\sigma_{i}))$
          \State Calculate policy loss $L(\theta)$ using $\hat{r}(s_{t},a_{t},\sigma_{i})$
          \State Update policy $\theta \xleftarrow[]{} \theta - \alpha \nabla_{\theta}L(\theta)$
      \EndFor
  \end{algorithmic}
  \end{algorithm}
    
Algorithm \autoref{alg:algorithm1} presents the TradeR algorithm corresponding to an RL agent. The hierarchical policy of the RL agent $\pi_{\theta}$ is composed of order and bid policies, $\pi_{ord}$ and $\pi_{bid}$, respectively and parameterized using $\theta$. We use $\theta$ to jointly denote the parameters of the order and bid policies. Corresponding to each step in the environment, the agent observes state $s_{t}$ as a vector of prices and volumes of symbol corresponding to the past 5 timesteps. The agent then samples an order quantity $y_{t}$ from its order policy $\pi_{ord}(\cdot|s_{t})$. Based on the order quantity, the agent samples the final bid $a_{t}$ from its high-level bid policy $\pi_{bid}(\cdot|s_{t},y_{t})$ which it executes in the environment to transition to next state $s_{t+1}$ and observe reward $r_{t}$. The replay buffer is updated with the corresponding $(s_{t},a_{t},r_{t},s_{t+1})$ tuple. 

During the learning phase, the agent samples a random batch of transitions from the replay buffer and computes standard deviations $\sigma_{i}$ between consecutive states of the batch. The surprise value function $V_{surp}$ estimates surprise based on standard deviation values $\sigma_{i}$ which is further encoded using the energy-based Mellowmax operator. The energy-based estimate is weighed using a temperature parameter $\beta$ with the transition reward $r(s_{t},a_{t})$ to yield the complete reward signal $\hat{r}(s_{t},a_{t},\sigma_{i})$. Finally, the agent's policy is updated using a standard RL loss $L(\theta)$ computed from reward $\hat{r}(s_{t},a_{t},\sigma_{i})$ and target estimates $Q(s_{t+1},a_{t+1})$.

\section{Experiments}
Our experiments aim to assess practical application of TradeR and its components. More specifically, we address the following three questions- 
\begin{itemize}
    \item How can TradeR be applied to practical trading pipelines consisting of real market data?
    \item What trends (if any) does TradeR present which would aid in understanding its practical utility?
    \item What contributions do TradeR's components hold?
\end{itemize}

\subsection{Learning Trade Execution}
We begin by analyzing the practical utility of TradeR from real market data. We collected price transitions from the Global US Stock Market for 2019 fiscal year comprising of COVID19 market crash. In order to tackle catastrophy and surprise minimization, we aim to challenge the algorithm with regard to the best and worst case scenarios. To this end, we shortlist 20 best and 15 worst stock symbols from the S\&P500 index consisting of price transitions at 1 minute intervals. TradeR is combined with Proximal Policy Optimization (PPO) \cite{ppo} and compared with strong baselines such as Deep Deterministic Policy Gradient (DDPG) \cite{ddpg} and Twin-Delayed DDPG (TD3) \cite{td3}. In the interest of evaluating data-efficiency and performance over longer temporal spans, agents were trained for only $500$ episodes consisting of $10^{4}$ steps each. Note that this is a significantly challenging training setup when compared to conventional RL benchmarks \cite{mujoco,dm,d4rl,meta} which train locomotion agents upto $10^{3}$ steps. In order to evaluate practical utility  of agents, we normalize all returns with respect to maximum profits earned by a professional trading algorithm. We direct the curious reader to \autoref{appendix:implem} for details on experimental setup.

\begin{figure*}[ht]
  \centering
  \begin{subfigure}[b]{0.4\textwidth}
      \centering
      \includegraphics[height=3.25cm,width=0.8\textwidth]{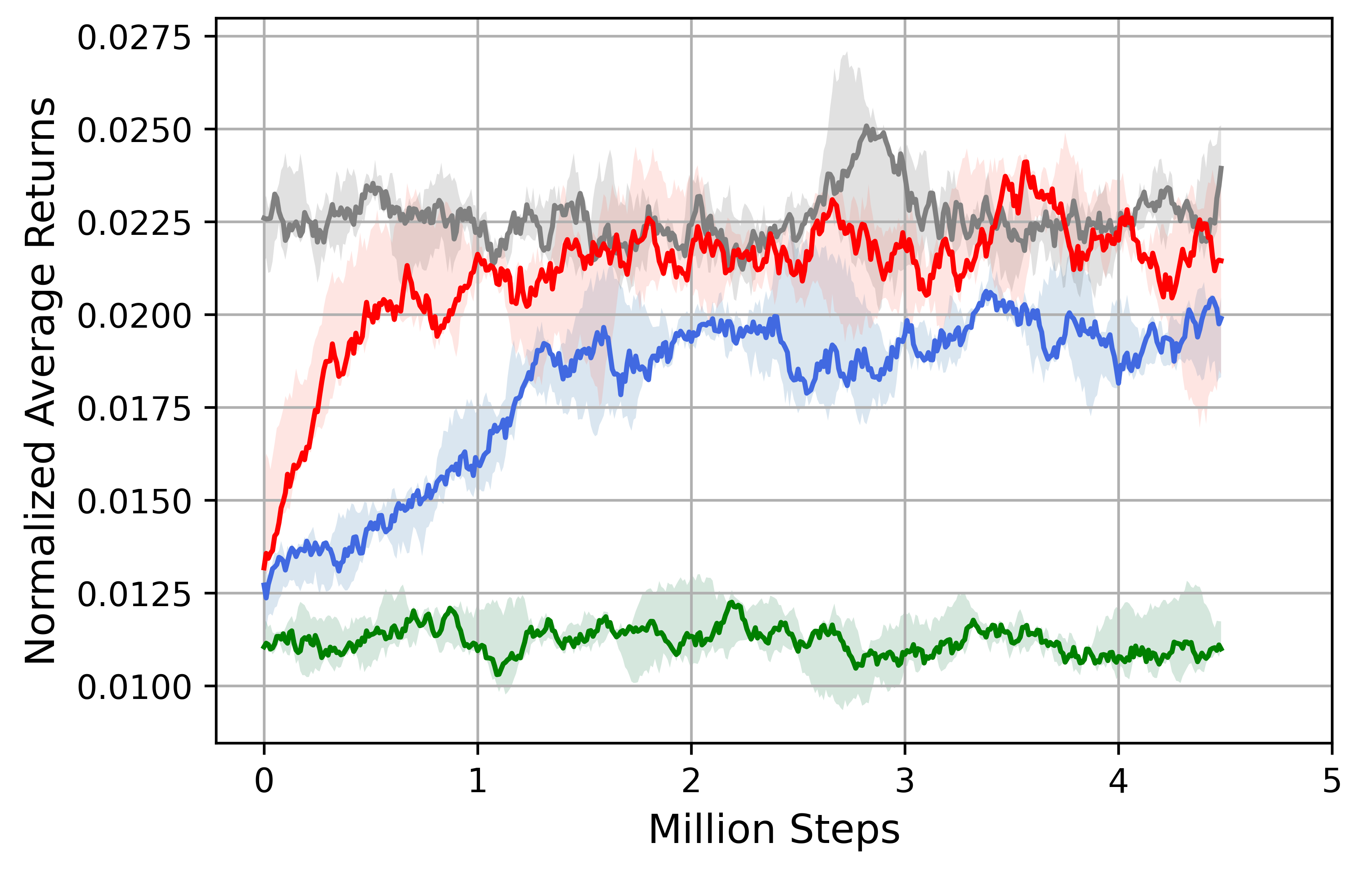}
  \end{subfigure}
  \begin{subfigure}[b]{0.4\textwidth}
      \centering
      \includegraphics[height=3.25cm,width=1.1\textwidth]{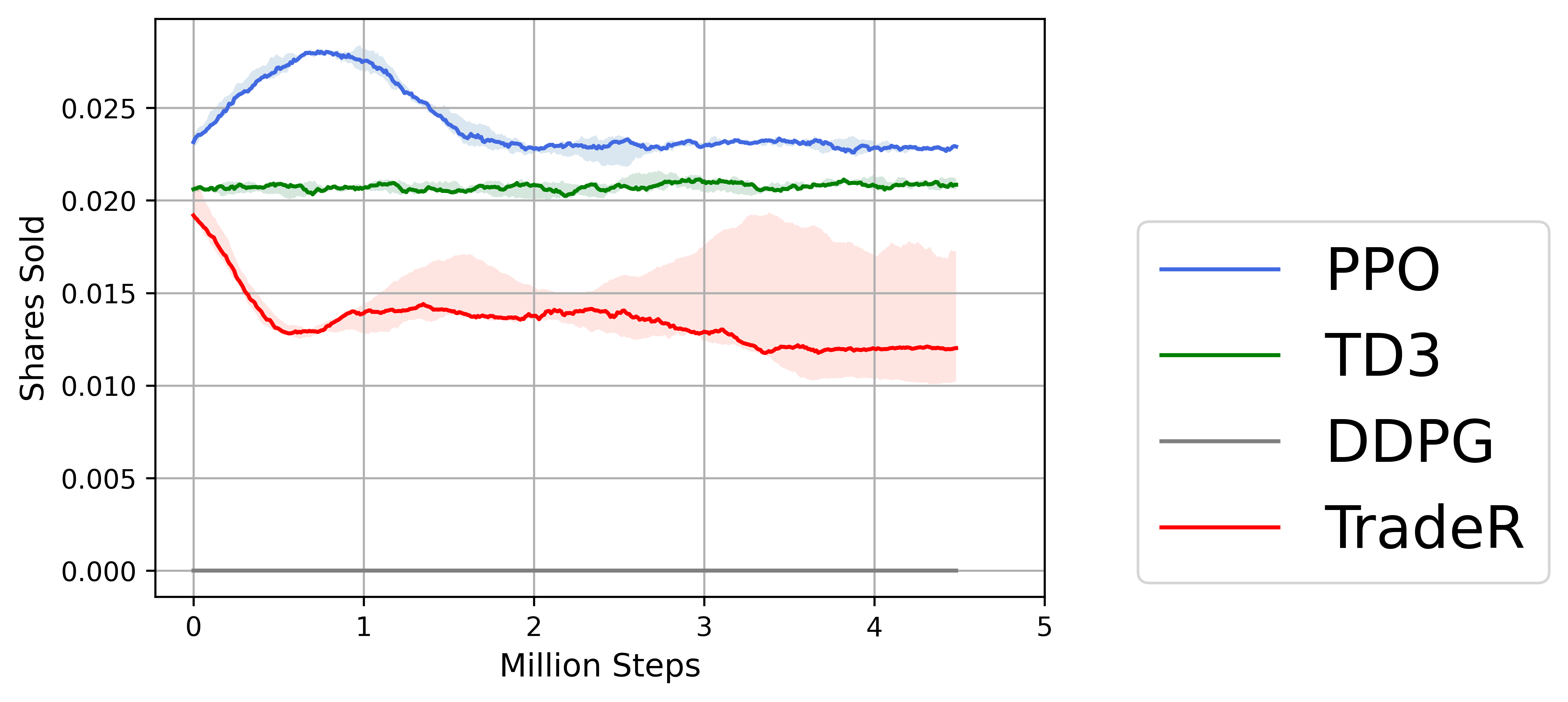}
  \end{subfigure}
     \caption{Left: Average normalized returns for all agents on the 35 symbol S\&P500 trading benchmark. TradeR demonstrates consistent gains over its (5) random seeds. Right: Average shares sold by agents during training. TradeR exhibits a safer strategy and holds on to its assets for long-term gains.}
     \label{fig:perf}
\end{figure*}

\autoref{fig:perf} (left) presents the combined normalized performance of TradeR on all symbols in comparison to baseline methods and PPO. TradeR learns quickly as a result of bi-level abstract hierarchies in the framework. Estimation of meaningful order quantities by the order policy provisions the high-level bid policy to learn suitable bidding strategies which lead to profitable episodes in the long-horizon. While deterministic execution algorithms such as DDPG and TD3 demonstrate high variance across random seeds, TradeR presents consistent trends in its returns.

\begin{figure*}[ht]
    \centering
    \includegraphics[height=4cm,width=16cm]{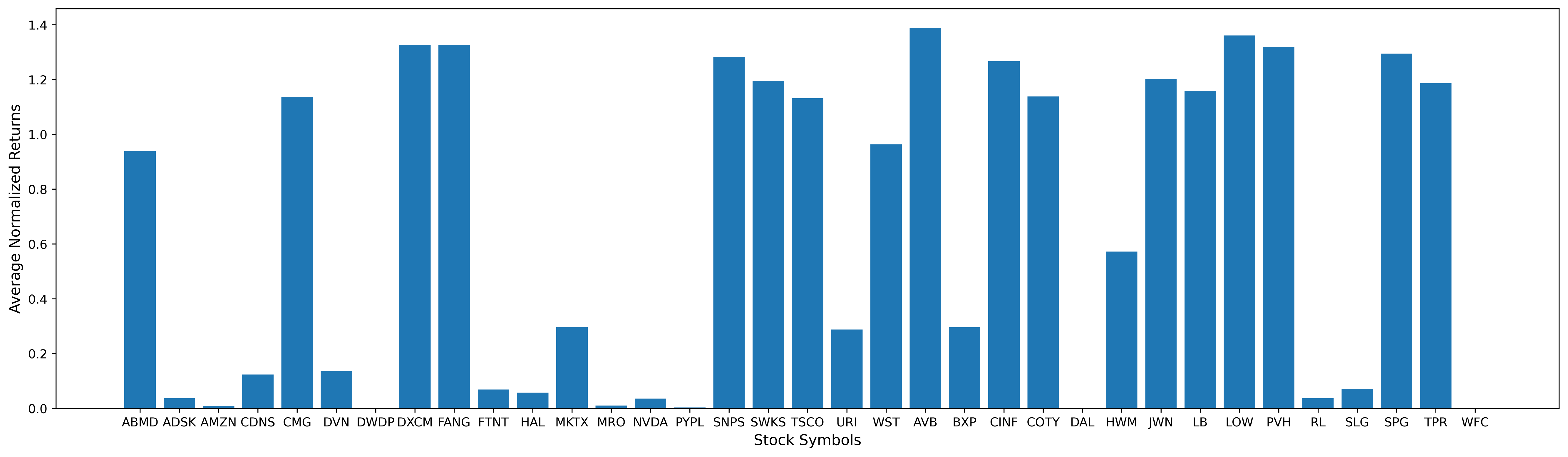}
    \caption{Performance of TradeR observed on all 35 S\&P500 symbols consisting of best (ABMD-WST) and worst (AVB-WFC) performing stocks for 2019 fiscal year. Rewards are normalized over a professional state-of-the-art trading algorithm.}
    \label{fig:all}
  \end{figure*}
  
While improved returns highlight the suitability of an RL algorithm, they do not necessarily throw light on its practical viability. Various RL methods demonstrate superhuman level performance \cite{dqn,ddqn} but only a handful can adapt towards real-world experience \cite{loon}. Motivated by this insight, we aim to answer the central insight of TradeR's practical utility. 

Towards this evaluation, we study the trading strategies learned by agents during training. Trading strategies are observed as per the manner in which an agent manages its share inventory. \autoref{fig:perf} (right) presents shares sold by agents across all symbols during training. TD3 and DDPG agents maintain deterministic policies and do not improve their trading strategies in accordance with policy improvement steps. PPO, on the other hand, presents a different trading strategy. The PPO agent begins by selling a large number of shares which is akin to huge losses for the 2019 fiscal year. Following the volatile trend observed in the market, policy improvement steps gradually improve the strategy of agent towards a conservative trading behavior. Upon approaching convergence, PPO reduces its share selling strategy and holds on to valuable assets with the motivation for long-term payoffs. 

Trading strategies for baseline methods fall prey to market crash and are only able recover after large number of updates (as in PPO). TradeR, on the other hand, exhibits a robust trading strategy which learns to minimize the number of shares sold efficiently. The order policy prefixes a low order quantity based on volatile price variations which the bid policy infers as a \textit{sell} bid. The TradeR agent thus continues to \textit{buy more} and \textit{sell less} leading to an optimal policy yielding high long-term returns. This key property of \textit{data-adaptive execution} is akin to \textit{catastophy} and \textit{surprise minimization}. 

To understand \textit{data-adaptive execution} we turn our attention back towards returns. As a consequence of best and worst symbols in the setup, TradeR demonstrates visible patterns which distinguish its performance across different trajectories. \autoref{fig:all} presents comparison of returns obtained by TradeR upon convergence on all 35 symbols in the setup. The arrangment (from left to right) depicts best (ABMD-WST) to worst (AVB-WFC) symbols during the pandemic. A notable finding one may observe is that TradeR performs consistently well on worst symbols with its performance variable for best ones. This stark distinction in TradeR's performance is reflective of its practical applicability highlighting greedy exploitation at lower prices. The agent familiarizes itself with the trend of price variations and utilizes the lower prices of worst symbols to maximize its payoffs. In the case of best symbols, the agent is aware of rising prices but at the same time conservative with regard to the volatility of the market. This results in a safe strategy reflecting the practical tradeoff between risk and profit in trading. 

\subsection{Ablation Study}
\begin{figure*}[ht]
    \centering
    \begin{subfigure}[b]{0.4\textwidth}
        \centering
        \includegraphics[height=4cm,width=0.9\textwidth]{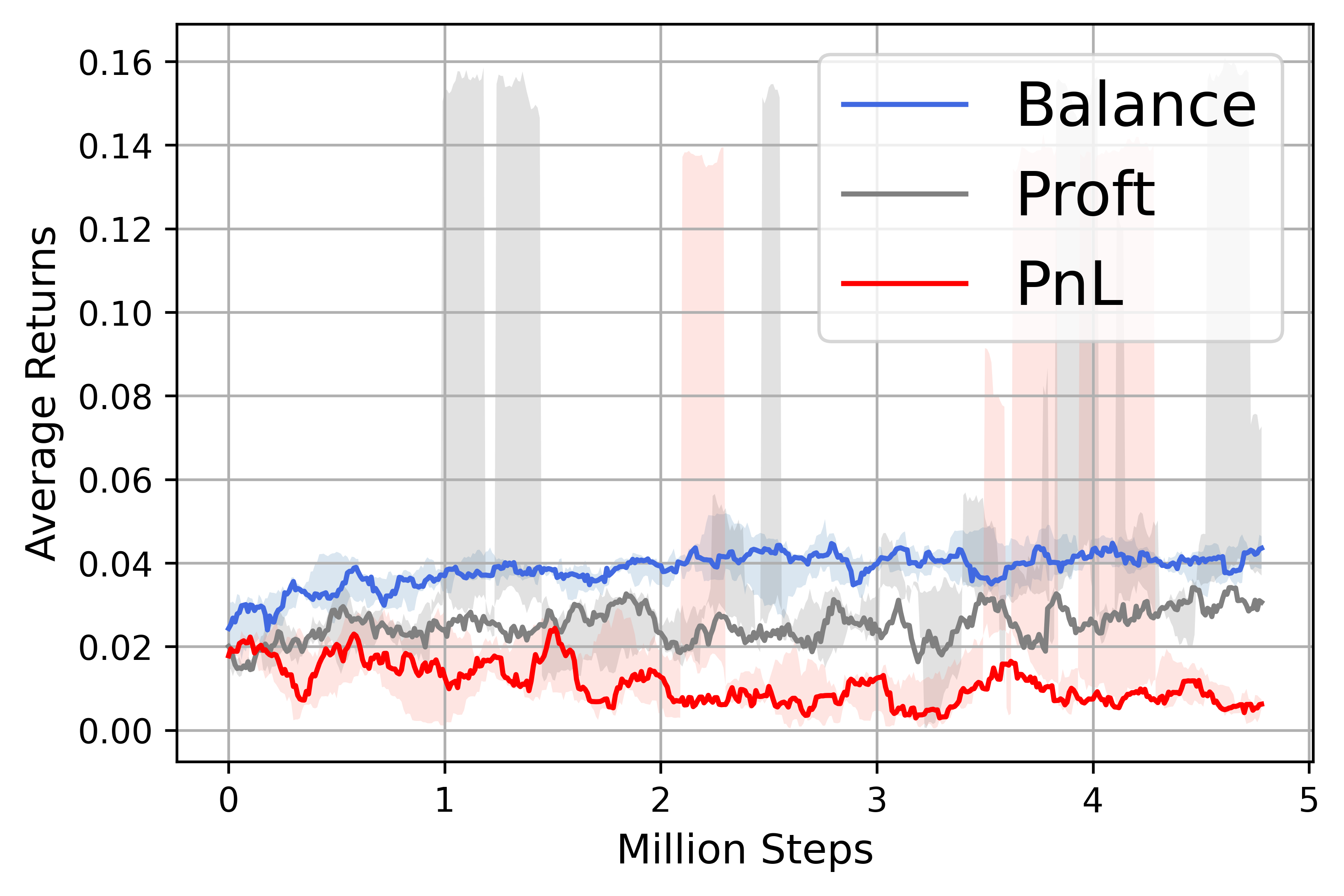}
    \end{subfigure}
    \begin{subfigure}[b]{0.4\textwidth}
        \centering
        \includegraphics[height=4cm,width=0.9\textwidth]{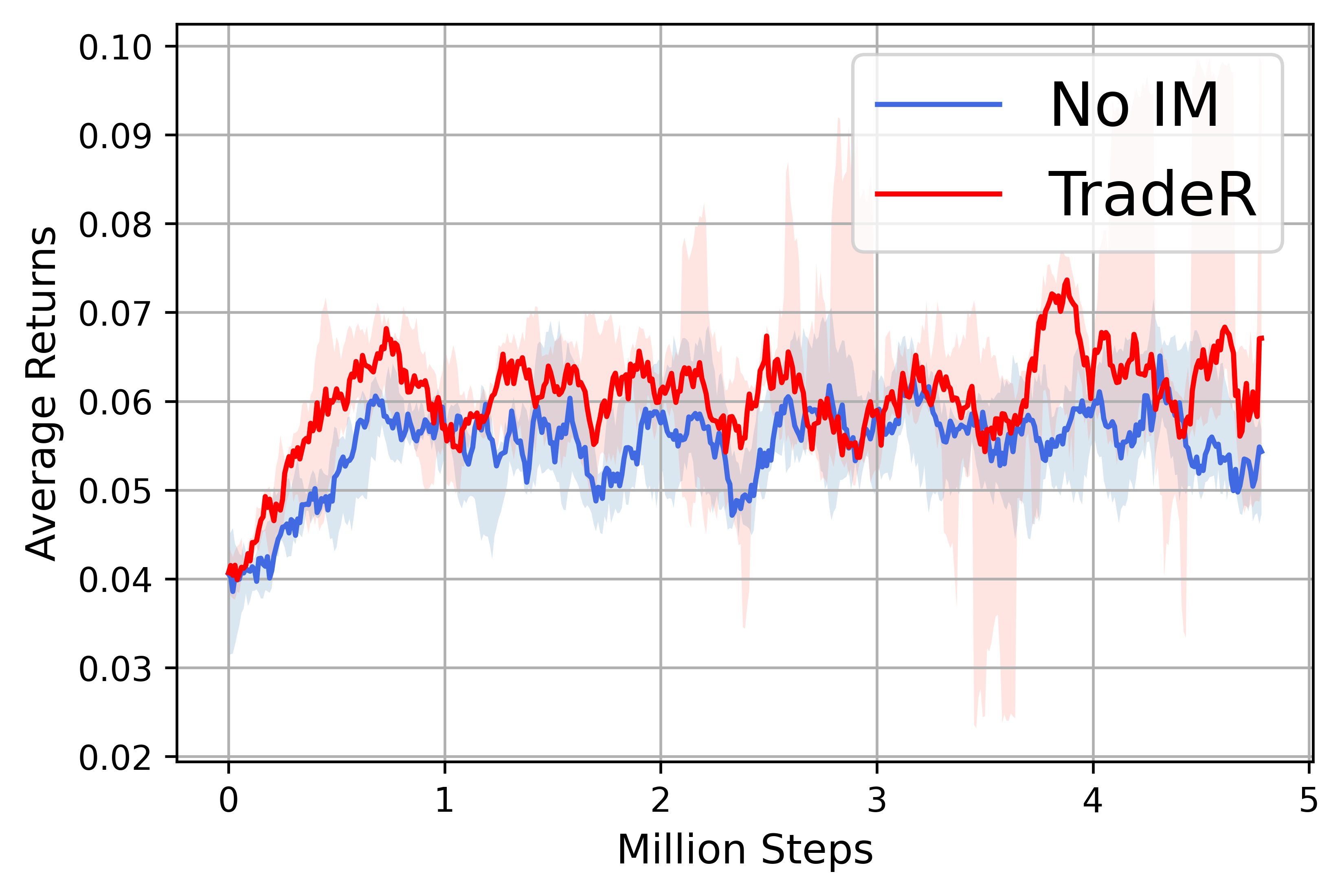}
    \end{subfigure}\\
    \begin{subfigure}[b]{0.4\textwidth}
      \centering
      \includegraphics[height=4cm,width=0.9\textwidth]{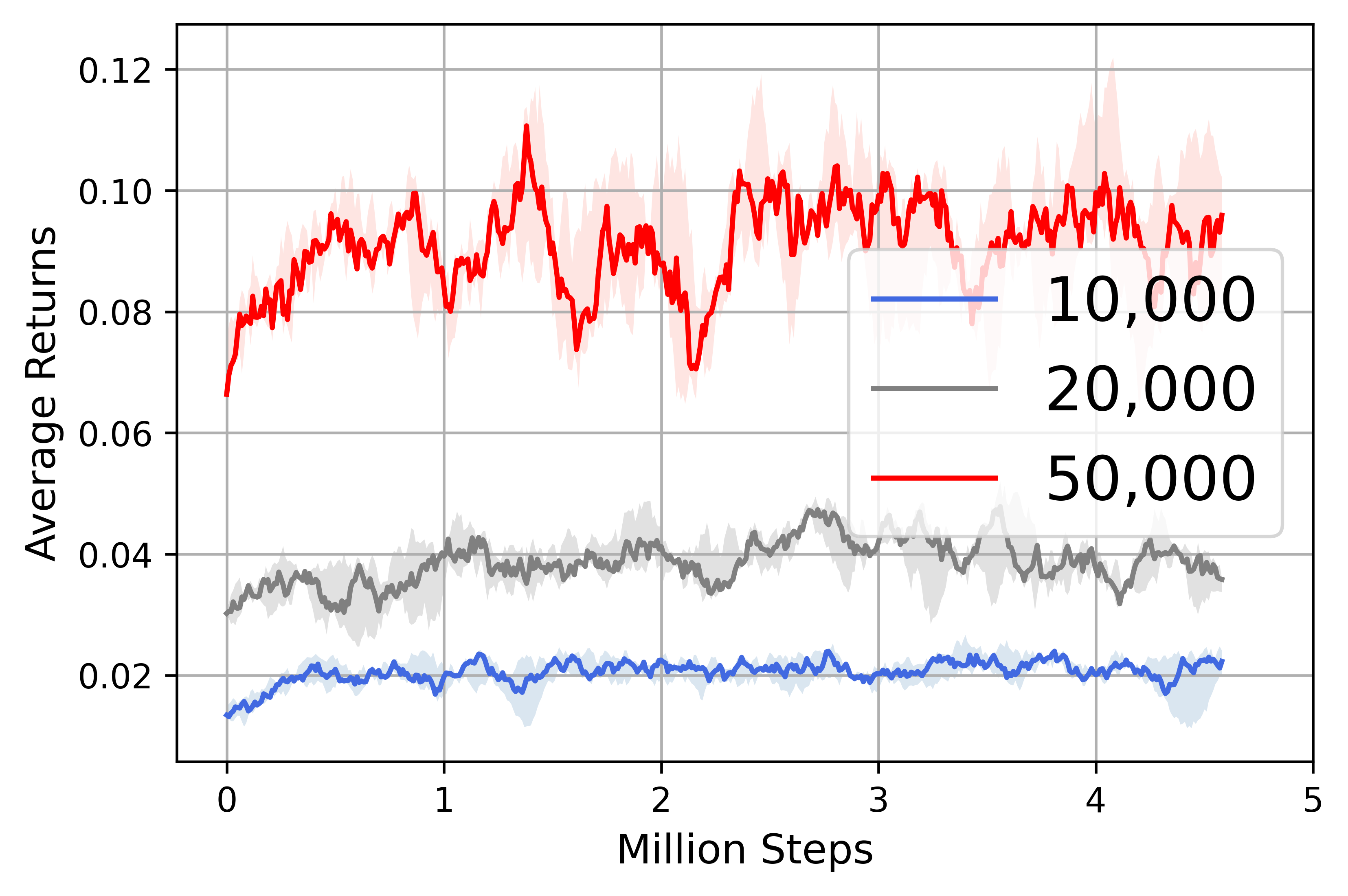}
  \end{subfigure} 
  \begin{subfigure}[b]{0.4\textwidth}
    \centering
    \includegraphics[height=4cm,width=0.9\textwidth]{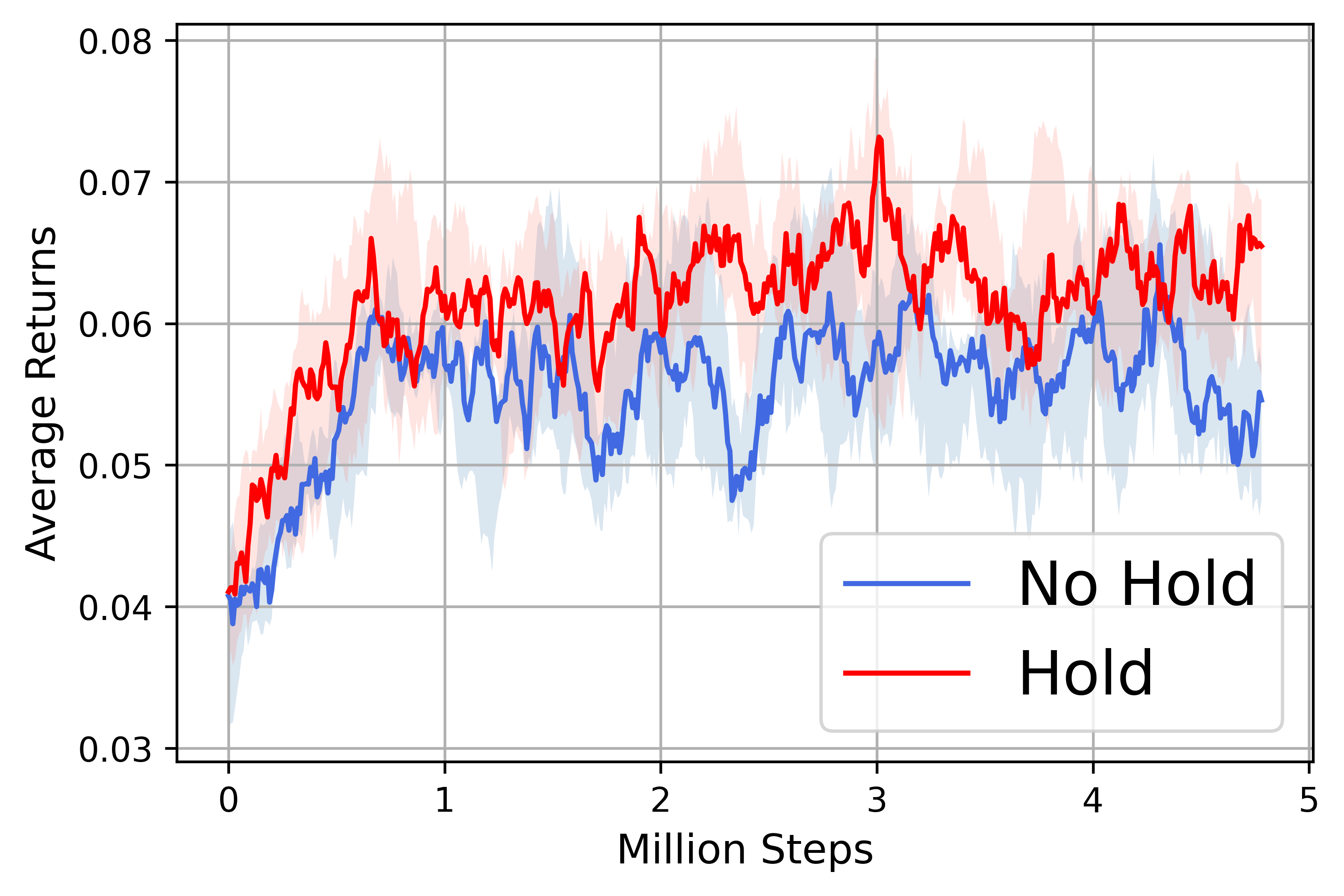}
  \end{subfigure} 
       \caption{Reward ablations for TradeR corresponding to all 35 symbols from the S\&P500 benchmark. Top-left: Balance amount serves as a suitable reward in comparison to per timestep high-variance profit and PnL.Top-right: Addition of Intrinsic Motivation (IM) facilitates surprise minimization. Bottom-left: TradeR demonstrates improving performance with increasing starting balance amounts. Bottom-right: Provision of hold action enables TradeR to learn apt long-horizon trading patterns.}
       \label{fig:ablations}
  \end{figure*}

We now evaluate efficacy of various components of TradeR framework. To this end, the ablation study also throws light on practical considerations such as the choice of reward function and the starting balance amount provided to the agent. Complete ablation study can be found in \autoref{appendix:ablations}.

\subsubsection{Reward Functions}
 \autoref{fig:ablations} (top-left) presents reward ablations carried out for different reward functions on the 35 symbol S\&P500 benchmark. We consider three reward choices, (1) \textit{balance amount} presents the current remaining balance of the agent, (2) \textit{Profits} indicate the net gains obtained from the start state of the episode until termination and (3) \textit{Profit and Loss (PnL)} which denotes the per-step gains/loss between consecutive states. PnL, being a per-step varying reward signal, presents high variance due to instantaneous profits and losses observed by the agent as a result of frequent price fluctuations. Since PnL is a noisy and transition-dependent reward signal, it hinders the agent from learning meaningful trading behaviors due to its strong dependence on price quotes. Profit, on the othert hand, does not provision stable learning as a result of invarying rewards observed from the 'hold' action. The agent often makes an early profit and does not act in subsequent timesteps leading to sub-optimal convergence. 
 
 Balance demonstrates an apt learning performance as the agent is incited to act out of the risk of going bust. From the perspective of practical trading scenarios, balance has another key advantage. In the case of most symbols, the balance reward metric presents the least deviation due to its reduced dependence on price fluctuations. This allows the agent to holistically make decisions based on volume and prices in contrast to solely weighing price quotes at the current timestep. 

\subsubsection{Energy-based Intrinsic Motivation}
\begin{figure*}[ht]
    \centering
    \begin{subfigure}[b]{0.4\textwidth}
        \centering
        \includegraphics[height=4cm,width=0.9\textwidth]{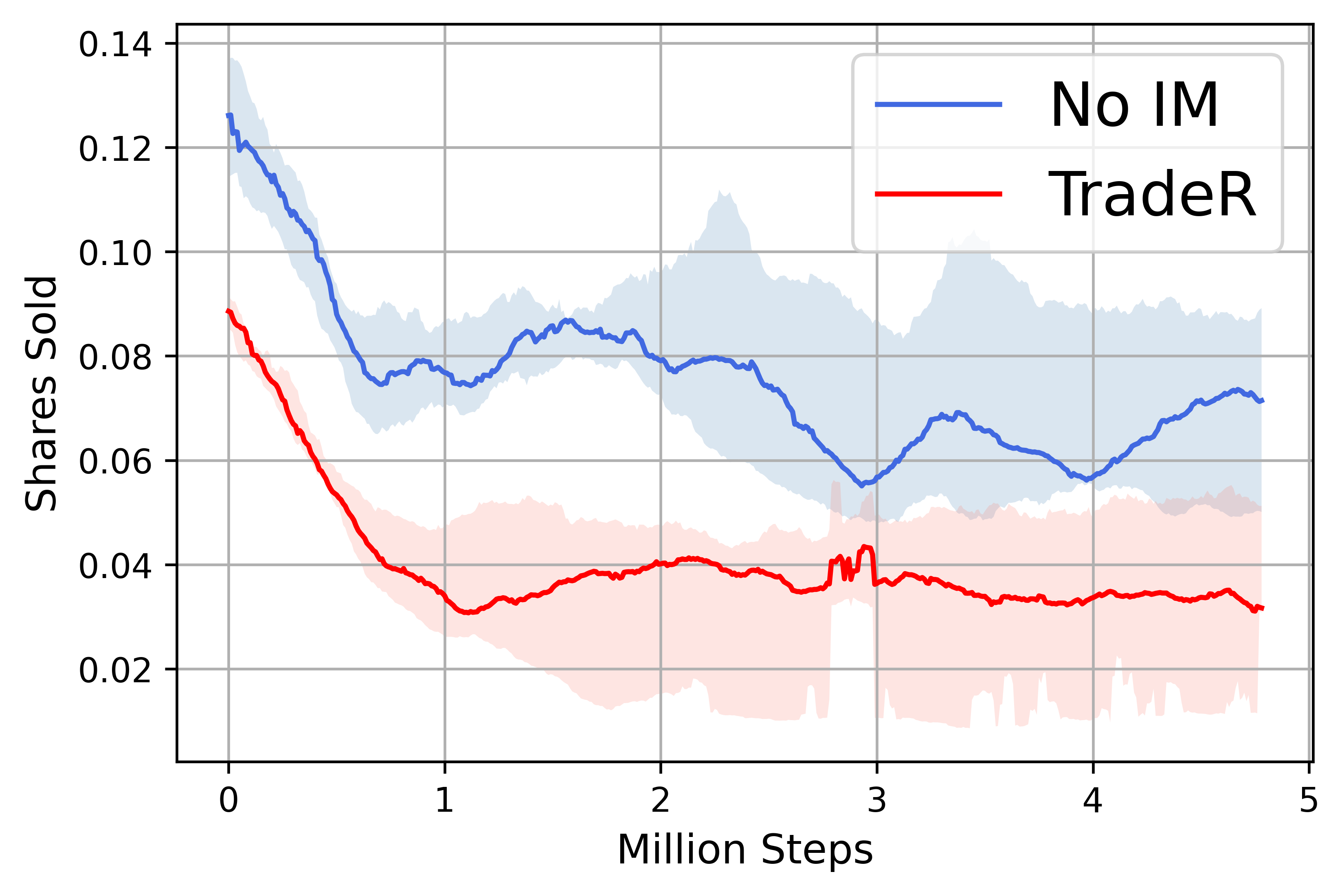}
    \end{subfigure}
    \begin{subfigure}[b]{0.4\textwidth}
        \centering
        \includegraphics[height=4cm,width=0.9\textwidth]{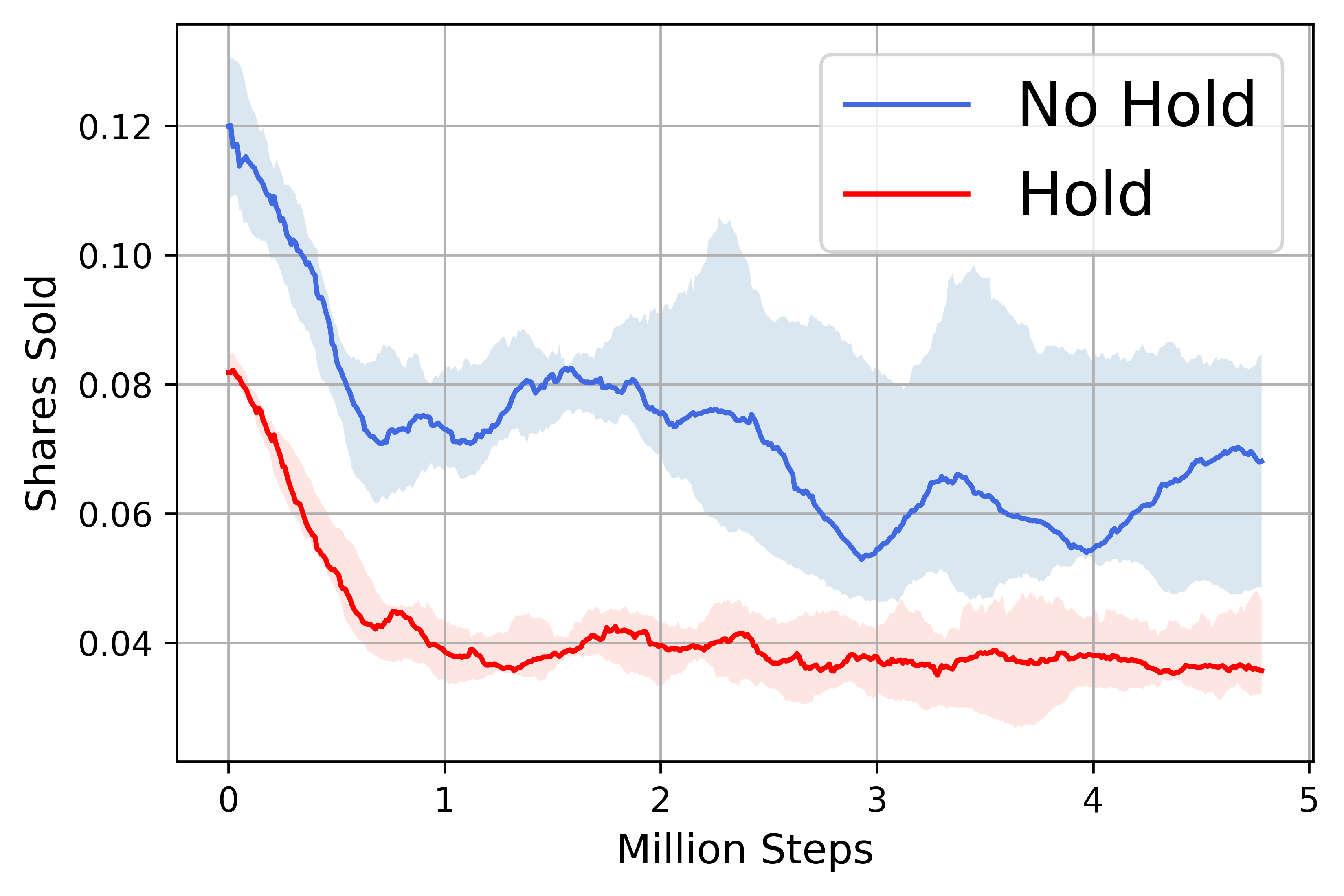}
    \end{subfigure}
    \caption{Ablations for total shares sold by TradeR agent over all 35 symbols from the S\&P500 benchmark. Left: The agent learns a \textit{buy more sell less} risk-averse strategy in the presence of Intrinsic Motivation (IM) which allows it to evade \textit{surprising} states. Right: Provision of hold action enables the agent to stabilize risk aversion and long-horizon gains which is otherwise found absent in prior trading mechanisms.}
       \label{fig:share_ablations}
  \end{figure*}
  
\autoref{fig:ablations} (top-right) presents reward ablations for the energy based scheme. In the absence of energy-based Intrinsic Motivation (IM), the agent learns slowly and demonstrates sub-optimal trading strategies which fail to retrieve profitable outcomes. Intrinsic motivation aids in efficient surprise minimization by evading catastrophic states as a result of safety-directed behavior. The agent, by virtue of estimated deviations in state-transition dynamic, is able to seek low energy configurations. This in turn induces sample efficient and stable learning over random seeds. 

\subsubsection{Starting Balance}
\autoref{fig:ablations} (bottom-right) presents reward ablations for different values of starting balance amounts. The TradeR agent is provided with $\$10K$, $\$20K$ and $\$50K$ as seed balance amount during training. The objective of this ablation experiment is to evaluate the stability of agent with growing amounts of trading burden. Performance of the framework is found consistently increasing with the increasing balance amount indicating its suitability to understand trade patterns and wisely leverage optimal policies for placing trade bids. Note that the agent does undergo catastrophic losses which is often the case with risk-taking human brokers who favour large bid amounts at larger balance values.

\subsubsection{Hold Action}
Contrary to pre-existing methods in literature \cite{ensemble,ddpo}, we provide the agent with an additional hold action which is equivalent to carrying out no transaction at the given timestep. \autoref{fig:ablations} (bottom-left) presents reward ablations for hold action. We compare the TradeR to the setting where the hold action is turned off and the agent is forced to trade by either buying or selling shares. TradeR demonstrated improved stable performance in the case of hold action indicating the necessity for a broader action scope which is often restricted in practical RL applications due to the curse of dimensionality. The hold action provides agent with the ability to simply save on previously bought earnings and channel a transaction only when the suitable price is encountered. This is in accordance with human patterns of trading wherein traders buy shares in excess and allow their savings to grow until a suitable quote is encountered.  

One can gain intuition into which components contribute the most towards TradeR's risk-averse and long-term behavior of \textit{buy more and sell less}. \autoref{fig:share_ablations} (left) presents ablations for shares sold by TradeR agent with and without the presence of energy-based intrinsic motivation scheme. TradeR demonstrates a lower number of total shares sold in the presence of intrinsic motivation validating its suitability for surprise and catastrophy minimization. Moreover, the agent presents a lower variance across its random seeds indicating consisting risk-averse behavior in conjunction with rewarding outcomes. 

Along similar lines, \autoref{fig:share_ablations} (right) presents ablations for shares sold by the TradeR agent with and without the provision of hold action. The agent favors holding values over longer time horizons and selling them once the suitable price quote is encountered. This allows the agent to sell more shares at lower prices rather than simply buying shares and accumulating assets in hopes for encountering a rewarding quote. Similar to \autoref{fig:ablations} (bottom-left), this behavior of agent is parallel to human brokers who prefer to wait for suitable quote and sell shares in large quantities.   

\section{Discussion}
Application of RL is hindered by practical challenges such as \textit{catastrophy} and \textit{surprise minimization}. In order to address these challenges, we focussed on the application of trading. We introduced the TradeR framework which is a hierarchical RL algorithm for practical high-frequency trading consisting of abrupt price variations. The lower level policy estimates order quantities on the basis of which the higher level policy executes trading bids in a data-driven market environment. TradeR utilizes an energy-based intrinsic motivation scheme in conjunction with surprise value function for estimating and minimizing surprise. Our large-scale study of 35 stock symbols obtained at 1 minute intervals from the S\&P500 index during 2019 fiscal year COVID19 market crash presents suitability of the hierarchical trading scheme. Furthermore, the ablation study conducted on payoffs and shares highlights the need for energy-based scheme in facilitating practical human-like trading patterns. 

Although TradeR presents a suitable mechanism for trading, the method suffers from two shortcomings. Firstly, the TradeR framework requires a large amount of data for training and evaluation on price-based experiences. While humans learn to trade intuitively with minimal (5-10 symbols) data requirements, TradeR requires large number of price transitions to understand evasion of surprising states. Secondly, transition data from the stock exchange is held static during TradeR's learning phase. This is not entirely reflective of an ever-moving market which consists of multiple brokers (\textit{agents}) continuously competing in the trading market (\textit{environment}). A more realistic setting would include trading as a multi-agent problem and tackle non-stationarity of dynamics. We aim to address these challenges for future work.

\bibliography{example_paper}
\bibliographystyle{icml2021}

\newpage
\appendix
\onecolumn    

\section{Implementation Details}
\label{appendix:implem}
This section highlights our setup for training and evaluating the TradeR framework. We combine TradeR with Proximal Policy Optimization (PPO) consisting of an Actor-Critic framework \cite{a2c,a3c}. The Actor consists of order and bid networks as two seperate policies. Order quantity $y_{t}$ estimated by the order network using policy $\pi_{ord}(y_{t}|s_{t})$ is concatenated with the environment state $s_{t}$ and provided as input state to the bid network. The bid network utilizes the high-level bid policy $\pi_{bid}(a_{t}|s_{t},y_{t})$ to estimate the final bid $a_{t}$ as environment action. The agent, consisting of order and bid policies, is trained jointly using policy gradient framework provisioned by the critic. During policy gradient updates, the surprise value function is utilized to estimate surprise using standard deviations across state dimensions which are then operated upon by the Mellowmax operator. An alternate method for applying the surprise value function would be to estimate surprise corresponding to each state. However, we empirically found this technique to demonstrate high variance in its estimates and often steer the agent towards surprising states. In order to further stabilize training, we add a small penalty to agent's loss function in the form of energy-based surprise estimates. This ensures that information about surprising states is propagated directly into actor's policies which may not be the case for some symbols in the training benchmark.  

\subsection{S\&P 500 Data}
Market data utilized for learning and evaluation of TradeR was collected from the S\&P500 market index. This stock market index is similar to other global market indices such as Dow and Jones 100 index. We specifically select S\&P500 as a benchmark as it cosists of a wider variety of stock symbols and suitably demonstrates global market trends. The data collected consists of price and volume transitions which are provided as state to the agent. More specifically, the following transitions were collected and presented to the agent at each timestep-

\textbf{Open Price:} The opening price of stock symbol in the market for current day.\\
\textbf{Close Price:} The closing price of stock symbol in the market for past day.\\
\textbf{Low Price:} The lowest price observed for the stock symbol in the market during current day.\\
\textbf{High Price:} The highest price observed for the stock symbol in the market during current day.\\
\textbf{Volume:} The quantity of shares which changed hands between the current and past timesteps.

All prices were observed in US Dollars (\$). Transitions were observed at 1 minute intervals from 35 stock symbols during the 2019 fiscal year ranging from $1^{st}$ April 2019 to $31^{st}$ March 2020. This includes the COVID19 market crash which was observed in February 2020. We select 35 stock symbols (20 best and 15 worst) from the market as these were reflective of majority of symbols in the market during the later half of 2019.

\subsection{Computational Requirements}

The agent was trained for a total of 5 million steps over 5 random runs for each symbol. This results in 6 GPU hours of training for a single run of one symbol. The total computational cost of training TradeR was 30 GPU hours for one symbol and 1050 GPU hours for all 35 symbols. In the interest of time, we parallelized computation on 4 NVIDIA RTX2080 Titan GPUs which reduced the time to 262.5 GPU hours resulting in 11 days of training. However, the major computational bottleneck resulted from the ablation study of TradeR. A total of 5 ablation experiments in addition to TradeR experiments were performed which increased the computation sixfold. Thus, the total computational time required for training and evaluating TradeR was $(6 \times 262.5 = 1575)$ GPU hours resulting in 66 days of experiments. Each training run consisted of 500 episodes of $10^{4}$ timesteps with policy gradient updates carried out at every 100 steps.  We tried longer timesteps of $12,000$ and $15,000$ but these did not demonstrate significant robustness to variance in agent's performance. In the case of $15,000$ timesteps, the agent often steered away from its consistent behavior and incurred lossy outcomes which were otherwise found absent in the $10^{4}$ timesteps setting. 

\subsection{Training Details}

The order network consists of an input layer of 128 units with tanh nonlinearity, a hidden layer of 128 units with tanh nonlinearity and an output layer of 1 unit with sigmoid nonlinearity. The bid network has a similar architecture except that the output layer has 3 units each corresponding to its 3 bid actions of buy, sell or hold. Action outputs from both networks are concatenated to yield the final 2-dimensional vector comprising of an order quantity in the range $[0,1]$ and a discrete action from the set $\{0,1,2\}$ denoting the type of bid. The architecture for critic network is same as that of order network. In the case of surprise value function, we adopt a different architecture. The two-layered network consists of an input layer of 128 units with ReLU \cite{relu} nonlinearity followed by an output layer having number of units equivalent to the number of state dimensions. We tried a range of different architectures for surprise value function such residual and skip connections. However, these did not demonstrate significant improvement. Finally, the TradeR agent was optimized using the Adam optimizer \cite{adam}. 

\begin{algorithm}[H]
  \caption{Trading Environment}
  \begin{algorithmic}[1]
    \label{alg:algorithm2}
    \State \textbf{Input: }$S,a_{t}$
    \State \textbf{Output: }$r_{t},s_{t+1}$
    \State Initialize account\_balance, max\_account\_balance, max\_num\_shares, max\_share\_price, max\_open\_positions
    \State Initialize net\_worth, shares\_held, shares\_sold, cost\_basis, sales\_value
    \State / / / / / / / / / / / / FETCH NEXT OBSERVATION FROM SEQUENCE
    \Function{observation}{$S$}
    \State $obs\xleftarrow[]{}\frac{S[-5:]}{max\_share\_price}$
    \State $obs \xleftarrow[]{} (obs|[balance,net\_worth, shares\_held, shares\_sold, cost\_basis, sales\_value])$
    \State \Return obs
    \EndFunction
    \State / / / / / / / / / / / / EXECUTE ACTION BY BUYING/SELLING SHARES
    \Function{step}{$S,a_{t}$}
      \If{$a_{t}==0$}:
          \State $shares\_bought \xleftarrow[]{} \frac{balance}{S[-1]} \times amount$
          \State $prev\_cost \xleftarrow[]{} cost\_basis \times shares\_held$
          \State $add\_cost \xleftarrow[]{} shares\_bought \times S[-1]$
          \State $balance \xleftarrow[]{} balance - add\_cost$
          \State $cost\_basis \xleftarrow[]{} \frac{prev\_cost + add\_cost}{shares\_held + shares\_bought}$
          \State $shares\_held \xleftarrow[]{} shares\_held + shares\_bought$
      \ElsIf{$a_{t}==1$}:
          \State $shares\_sold \xleftarrow[]{} shares\_held \times amount$
          \State $balance \xleftarrow[]{} balance + shares\_sold\times current\_price$
          \State $shares\_held \xleftarrow[]{} shares\_held - shares\_sold$
          \State $sales\_value \xleftarrow[]{} shares\_sold \times current\_price$
      \EndIf
      \State $net\_worth \xleftarrow[]{} balance + shares\_held \times current\_price$
      \State $reward \xleftarrow[]{} balance$
      \State $obs \xleftarrow[]{} OBSERVATION(S)$
      \State \Return reward, obs
    \EndFunction
    \State / / / / / / / / / / / / RESET ENVIRONMENT TO A RANDOM STATE
    \Function{reset}{S}
      \State $balance \xleftarrow[]{} account\_balance$
      \State $net\_worth \xleftarrow[]{} account\_balance$
      \State $shares\_held \xleftarrow[]{} 0$
      \State $shares\_sold \xleftarrow[]{} 0$
      \State $sales\_value \xleftarrow[]{} 0$
      \State $cost\_basis \xleftarrow[]{} 0$
      \State $obs \xleftarrow[]{} observation(S)$
      \State \Return obs
    \EndFunction
  \end{algorithmic}
\end{algorithm}

\subsection{Execution Setup}
The agent interacts in a custom-designed trading environment which replicates the buying and selling of shares in real global markets. The motivation behind the environment stems from Capital Markets framework which consists of various different order bids being processed at the same time. \autoref{fig:TradeR} presents the dynamic trading environment used to simulate real-world market data. The price sequence corresponding to a symbol is obtained from the stock exchange which is stored in a prefetch buffer. Similar to price quotes, the symbol metadata (date-time intervals and volume transitions) are appended to the sequence variable in the prefetch buffer. A batch of transitions is sampled from the prefetch buffer and preprocessed using the transition preprocessor. Preprocessing steps on a batch include filtering and sorting of price quantities and volume vectors according to their date-time interval. The transition preprocessor feeds the processed transitions into a transition queue which preserves the temporal structure of prices as in the real stock market. Each entry in queue is sampled for  normalization and scaling as per the maximum quantity in price vectors. This is an additional step in comparison to the real market environment in order to facilitate accurate approximations and lower variance in order quantities of the agent. Normalized and scaled transitions are passed to the action simulator which simulates the transitions using internal local variables based on actions selected by the agent. At each timestep, the action simulator queries the reward function and yields the next state $s_{t+1}$ and reward $r_{t}$ to the agent. An additional technique to speed up learning would be to make use of multiple action simulators in an asynchronous fashion which obeys the real market trading constraints. However, this requires multiple actions to be queried by the agent while updating its policy resulting in large computational requirements. 

Algorithm 2 presents the trading setup used for simulating a real-world market scenario. Stock sequence $S$ corresponding to a symbol is traversed on the basis of shares bought and sold during agent's execution. Following are the three key functions-\\
RESET- This method resets the environment to an initial random state with all data variables initialized to 0. \\
OBSERVATION- This method processes the observation variables $obs$ by combining micro-level observations such as price quotes with macro-level observations such as the current balance, net worth, total shares held and sold by the agent. The combined observation is $36$-dimensional vector which represents the state of the agent. \\
STEP- The STEP method transitions the agent from current to next timestep. Based on the action $a_{t}$ selected by the agent, the function updates macro-level observations which are stored in a queue-like fashion. Following macro-level update of dynamics, the method provides a scalar reward value proportional to the change in prices which leads to the next-state obtained using OBSERVATION method.

\subsection{Note on Comparison with Soft Actor Critic (SAC)}
In orde to assess the efficacy of TradeR with state-of-the-art RL methods, we tried to implement SAC on our real-market trading environment. However, SAC presented significantly high variance in its returns with no meaningful trading strategies learned over the course of training. To address this issue, the entropy parameter of SAC was tuned for 5 different values between 0 and 0.5 in steps of 0.05. While lower values of entropy parameter demonstrated learning of algorithm, the returns obtained by SAC agent were not comparable to any of the baseline agents and not consistent across its random seeds. We conjecture that SAC does not perform well under abrupt dynamics as significant deviations between consecutive states hurt the entropy of action distribution. We leave the comparison of TradeR with SAC as a potential direction for future work. 

\subsection{Hyperparameters}
\autoref{tab:hyp} presents the hyperparameters used for training the PPO agent using TradeR framework. Learning rate $\alpha$ was tuned for values $0.0001, 0.0003$ and $0.0005$ with $0.0003$ being the suitable one. Energy temperature $\beta$ was tuned for five different values $0.001,0.005,0.01,0.05$ and $0.1$. Lower values presented stability but slower learning with $\beta = 0.01$ demonstrating stable and fast learning of surprising states. Lastly, policy update duration was set to $100$ steps after tuning it for $20,50,100,200,500$ and $1000$ steps where large values presented slower and sub-optimal convergence.
\begin{table}[H]
  \centering
  \begin{tabular}{c|c}
       Hyperparameters & Values \\
       \hline
       timesteps per episode & $10000$ \\
       past prices as observations & $5$ prices\\
       batch size & $b=32$ \\
       learning rate & $\alpha=0.0003$ \\
       discount factor & $\gamma=0.99$ \\
       policy update interval & $100$ steps \\
       agent network hidden sizes & 128 \\
       surprise temperature & $\beta=0.01$
  \end{tabular}
  \caption{Hyperparameter values for PPO agent used with TradeR framework}
  \label{tab:hyp}
\end{table}

\section{Complete Results}

\begin{table}[H]
  \centering
  \begin{tabular}{l|l|l|l|l}
  Symbol & TradeR    & PPO       & TD3       & DDPG       \\
  \hline
  ABMD   & \textbf{0.94±0.04} & 0.76±0.08 & 0.44±0.03 & 0.87±0.06  \\
  ADSK   & \textbf{0.04±0.01} & 0.03±0.03  & 0.02±0.03  & 0.04±0.04   \\
  AMZN   & 0.01±0.00  & 0.01±0.01 & 0.0±0.00   & \textbf{0.1±0.00}    \\
  CDNS   & \textbf{0.12±0.01} & 0.12±0.03 & 0.07±0.01 & 0.11±0.01  \\
  CMG    & \textbf{1.14±0.1}  & 0.69±0.19 & 0.23±0.02 & 0.44±0.02  \\
  DVN    & 0.14±0.07 & 0.12±0.02 & 0.08±0.01 & 0.\textbf{16±0.01}  \\
  DWDP   & \textbf{0.01±0.0}   & 0.0±0.0   & 0.0±0.0   & 0.0±0.0    \\
  DXCM   & \textbf{1.33±0.17} & 1.09±0.14 & 0.68±0.07 & 1.26±0.13  \\
  FANG   & \textbf{1.33±0.82} & 1.07±0.18 & 0.67±0.09 & 1.32±0.17  \\
  FTNT   & 0.07±0.01 & 0.06±0.0  & 0.04±0.0  & 0.\textbf{07±0.01}  \\
  HAL    & 0.06±0.0  & 0.06±0.01 & 0.04±0.0  & 0.\textbf{07±0.01}  \\
  MKTX   & \textbf{0.3±0.01}  & 0.19±0.01 & 0.12±0.0   & 0.18±0.01   \\
  MRO    & \textbf{0.01±0.01}  & 0.01±0.04  & 0.01±0.03  & 0.01±0.05   \\
  NVDA   & \textbf{0.04±0.0}  & 0.03±0.0  & 0.02±0.0  & 0.03±0.0   \\
  PYPL   & 0.0±0.0   & 0.0±0.0   & 0.0±0.0   & 0.0±0.0    \\
  SNPS   & \textbf{1.28±0.15} & 1.08±0.06 & 0.63±0.09 & 1.27±0.11  \\
  SWKS   & 1.2±0.17  & 1.15±0.03 & 0.65±0.07 & 1.\textbf{34±0.16}  \\
  TSCO   & 1.13±0.08 & 1.03±0.11 & 0.62±0.07 & 1.\textbf{19±0.15}  \\
  URI    & \textbf{0.29±0.04} & 0.25±0.02 & 0.15±0.02 & 0.27±0.04  \\
  WST    & \textbf{0.96±0.08} & 0.82±0.07 & 0.49±0.05 & 0.94±0.15   \\
  AVB    & \textbf{1.39±0.07} & 1.09±0.09 & 0.65±0.05 & 1.27±0.67  \\
  BXP    & \textbf{0.3±0.02}  & 0.25±0.02 & 0.15±0.01 & 0.15±0.16  \\
  CINF   & \textbf{1.27±0.75} & 1.14±0.1  & 0.69±0.07 & 1.23±0.72  \\
  COTY   & 1.14±0.07 & 1.16±0.18 & 0.71±0.07 & \textbf{1.28±0.74}  \\
  DAL    & 0.0±0.0   & 0.0±0.0   & 0.0±0.0   & 0.0±0.0    \\
  HWM    & 0.57±1.43 & 0.58±0.01 & 0.34±0.01 & 0.\textbf{67±0.33}  \\
  JWN    & 1.2±0.24  & 1.19±0.08 & 0.77±0.11 & \textbf{1.39±0.78}  \\
  LB     & 1.16±0.16 & \textbf{1.23±0.15} & 0.71±0.11 & 1.21±0.74  \\
  LOW    & \textbf{1.36±0.19} & 1.2±0.15  & 0.74±0.06 & 1.34±0.7   \\
  PVH    & \textbf{1.32±0.12} & 1.25±0.23 & 0.78±0.06 & 1.02±0.68  \\
  RL     & \textbf{0.04±0.01}  & 0.04±0.04  & 0.02±0.0  & 0.03±0.02  \\
  SLG    & 0.07±0.01 & 0.07±0.01 & 0.04±0.0  & 0.\textbf{08±0.04}  \\
  SPG    & \textbf{1.29±0.21} & 1.17±0.16 & 0.67±0.14 & 1.28±0.64  \\
  TPR    & \textbf{1.19±0.11} & 1.14±0.12 & 0.69±0.1  & 1.17±0.73  \\
  WFC    & 0.0±0.0   & 0.0±0.0   & 0.0±0.0   & 0.0±0.0   
  \end{tabular}
  \caption{Normalized average rewards for all 35 symbols from the S\&P500 benchmark observed at 1 minute intervals for 2019 fiscal year. Best peroformance corresponding to each symbol is highlighted in bold. TradeR demonstrates improved performance on 21 out of 35 symbols.}
\end{table}

\begin{figure}[H]
  \centering
  \includegraphics[height=22cm,width=17.5cm]{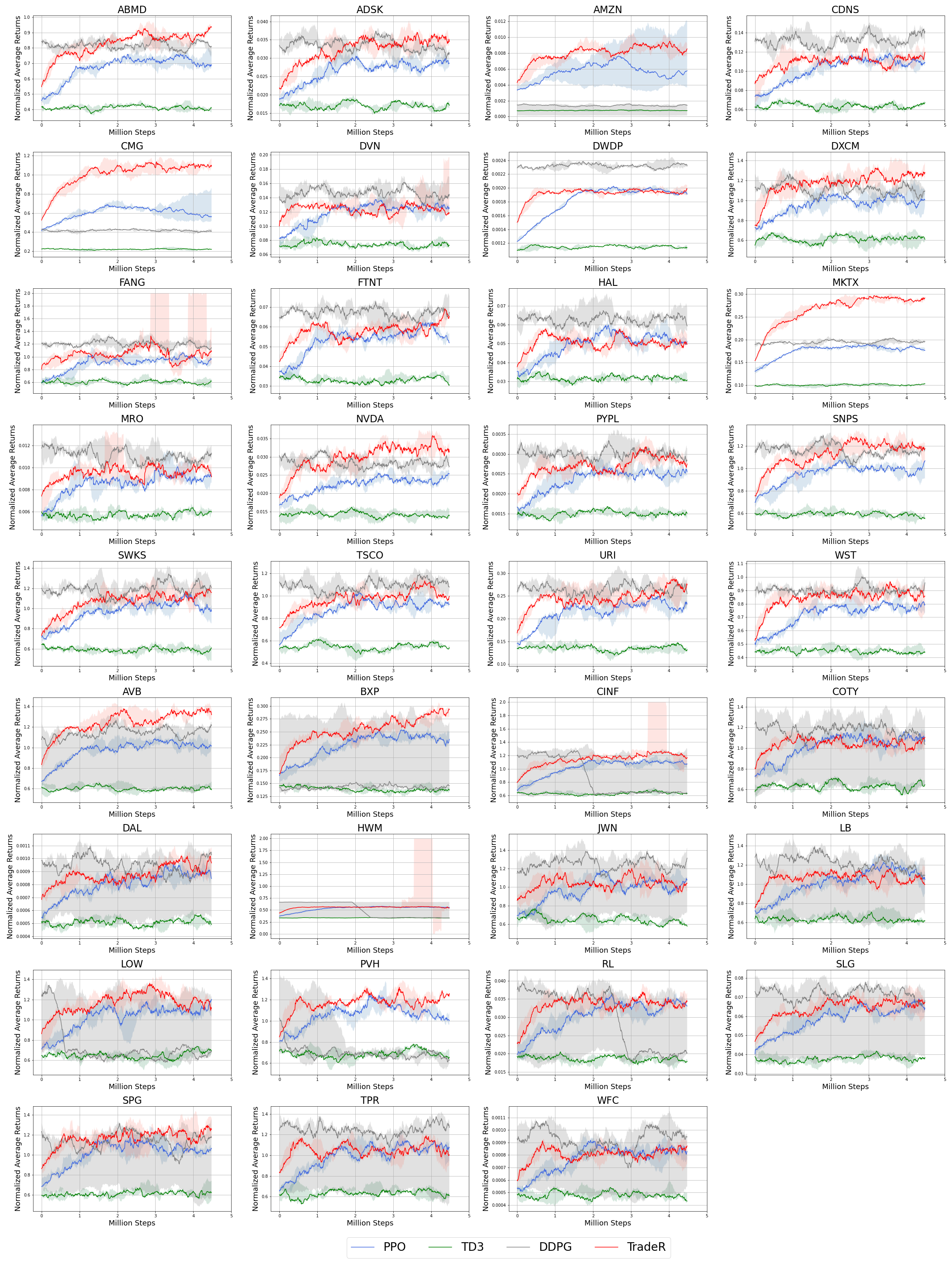}
  \caption{Normalized average rewards learned over 5 million steps for all 35 symbols. TradeR demonstrates consistent and sample-efficient performance across its 5 random seeds.}
\end{figure}

\begin{table}[H]
  \centering
  \begin{tabular}{l|l|l|l|l}
  Symbol & TradeR    & PPO       & TD3       & DDPG       \\
  \hline
  ABMD   & \textbf{0.61±0.05} & 0.87±0.04 & 0.63±0.02 & 0.0±0.0    \\
  ADSK   & \textbf{1.41±0.09} & 1.96±0.04 & 1.49±0.04 & 0.0±0.0    \\
  AMZN   & 0.23±0.03 & 0.4±0.21  & \textbf{0.03±0.0}  & 0.0±0.0    \\
  CDNS   & \textbf{0.3±0.03}  & 0.41±0.01 & 0.32±0.0  & 0.0±0.0    \\
  CMG    & 0.28±0.07 & 0.41±0.09 & \textit{0.15±0.0}  & 0.0±0.0    \\
  DVN    & \textbf{0.18±0.01} & 0.28±0.04 & 0.23±0.02 & 0.0±0.0    \\
  DWDP   & 0.0±0.0   & 0.0±0.0   & 0.0±0.0   & 0.0±0.0    \\
  DXCM   & \textbf{1.74±0.11} & 2.51±0.08 & 1.94±0.16 & 0.0±0.0    \\
  FANG   & \textbf{1.91±0.42} & 2.63±0.45 & 2.18±0.24 & 0.0±0.0    \\
  FTNT   & \textbf{1.66±0.17} & 2.39±0.06 & 1.86±0.04 & 0.0±0.0    \\
  HAL    & \textbf{1.34±0.33} & 1.88±0.24 & 1.58±0.23 & 0.0±0.0    \\
  MKTX   & 1.77±0.12 & 2.49±0.08 & \textbf{1.57±0.02} & 0.0±0.0    \\
  MRO    & \textbf{0.82±0.2}  & 1.28±0.32 & 1.03±0.17 & 0.0±0.0    \\
  NVDA   & \textbf{0.07±0.0}  & 0.1±0.0   & 0.07±0.01 & 0.0±0.0    \\
  PYPL   & 0.0±0.0   & 0.0±0.0   & 0.0±0.0   & 0.0±0.0    \\
  SNPS   & \textbf{1.66±0.13} & 2.37±0.1  & 1.8±0.03  & 0.0±0.0    \\
  SWKS   & \textbf{1.77±0.13} & 2.44±0.07 & 1.91±0.08 & 0.0±0.0    \\
  TSCO   & \textbf{1.59±0.16} & 2.32±0.04 & 1.8±0.04  & 0.0±0.0    \\
  URI    & \textbf{0.33±0.03} & 0.45±0.04 & 0.36±0.02 & 0.0±0.0    \\
  WST    & \textbf{0.77±0.08} & 1.12±0.01 & 0.85±0.03 & 0.0±0.0    \\
  AVB    & 1.66±0.33 & 2.23±0.07 & \textbf{1.64±0.04} & 0.0±1.66   \\
  BXP    & \textbf{1.27±0.11} & 1.79±0.08 & 1.36±0.01 & 1.33±1.39  \\
  CINF   & \textbf{1.37±0.09} & 2.02±0.05 & 1.61±0.07 & 1.59±1.6   \\
  COTY   & \textbf{1.45±0.21} & 2.07±0.26 & 1.76±0.14 & 0.0±1.64   \\
  DAL    & 0.0±0.0   & 0.0±0.0   & 0.0±0.0   & 0.0±0.0    \\
  HWM    & \textbf{1.42±0.05} & 2.09±0.05 & 1.65±0.03 & 1.66±1.66  \\
  JWN    & \textbf{1.54±0.03} & 2.16±0.23 & 1.74±0.12 & 0.0±1.75   \\
  LB     & \textbf{1.51±0.11} & 2.07±0.09 & 1.65±0.14 & 0.0±1.75   \\
  LOW    & \textbf{1.39±0.28} & 1.93±0.04 & 1.48±0.02 & 1.49±1.47  \\
  PVH    & \textbf{1.32±0.3}  & 1.98±0.22 & 1.54±0.21 & 1.53±0.22  \\
  RL     & \textbf{0.19±0.03} & 0.26±0.01 & 0.2±0.0   & 0.2±0.2    \\
  SLG    & \textbf{1.56±0.17} & 2.14±0.27 & 1.71±0.06 & 0.0±1.74   \\
  SPG    & \textbf{1.56±0.15} & 2.21±0.14 & 1.7±0.25  & 0.0±1.82   \\
  TPR    & \textbf{1.52±0.08} & 2.0±0.06  & 1.65±0.15 & 0.0±1.65   \\
  WFC    & 0.0±0.0   & 0.0±0.0   & 0.0±0.0   & 0.0±0.0
  \end{tabular}
  \caption{Normalized average shares sold for all 35 symbols from the S\&P500 benchmark observed at 1 minute intervals for 2019 fiscal year. Best peroformance corresponding to each symbol is highlighted in bold. While DDPG fails to learn a strategy for selling shares, TradeR demonstrates the human-like strategy of \textit{buy more sell less} on 27 out of 35 tasks.}
\end{table}

\begin{figure}[H]
  \centering
  \includegraphics[height=22cm,width=17.5cm]{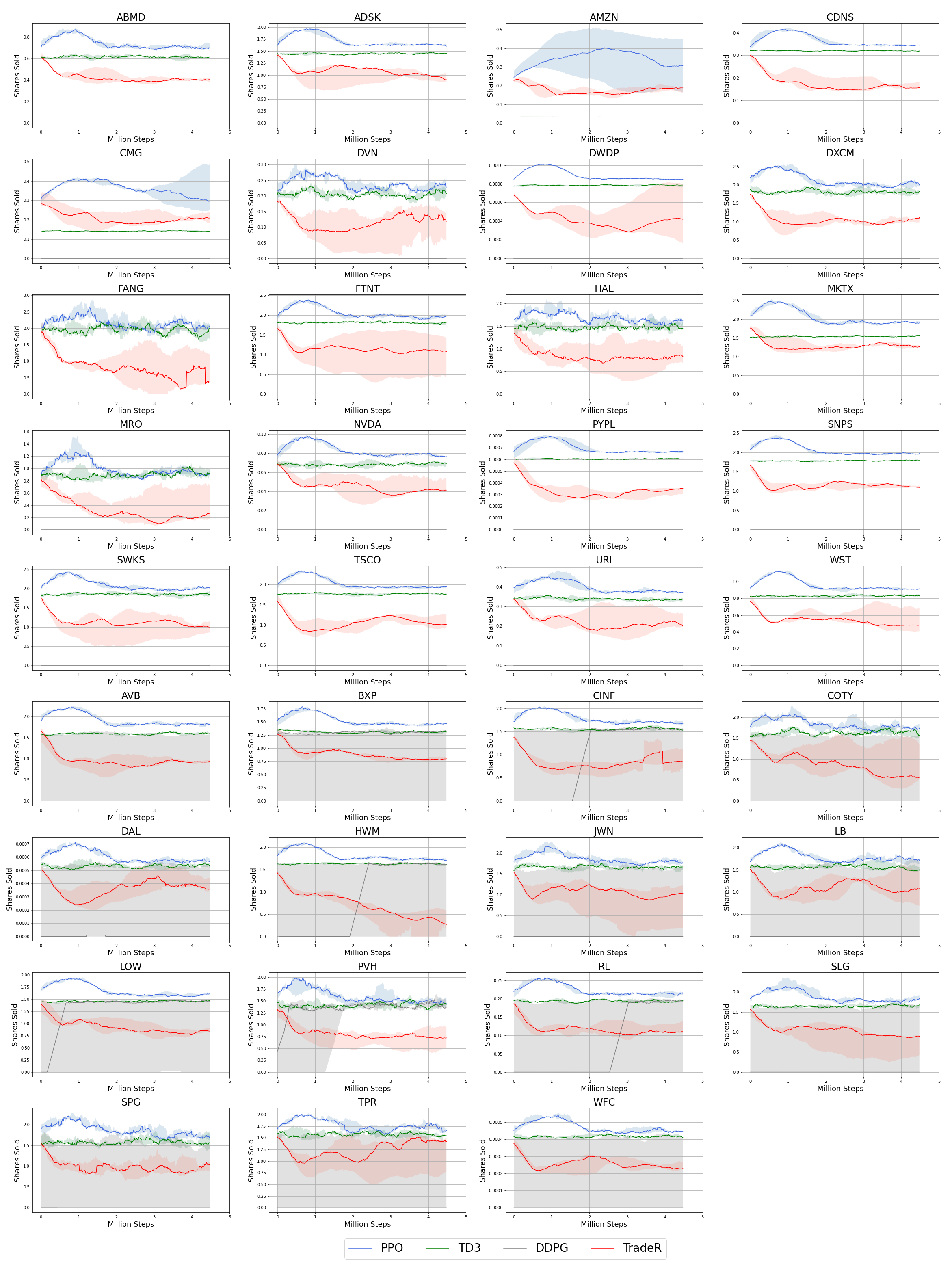}
  \caption{Normalized average rewards learned over 5 million steps for all 35 symbols. TradeR demonstrates consistent and sample-efficient performance across its 5 random seeds.}
\end{figure}

\section{Complete Ablation Study}
\label{appendix:ablations}
\subsection{Reward Functions}

\begin{table}[H]
  \centering
  \begin{tabular}{l|l|l|l}
    Symbol & Balance   & Profit    & PnL        \\
    \hline
    ABMD   & \textbf{1.43±0.19} & 0.82±0.29 & 0.81±3.34  \\
    ADSK   & 0.0±0.0   & 0.0±0.0   & 0.0±-4.0   \\
    AMZN   & 0.0±0.0   & 0.0±-4.0  & \textbf{0.59±-4.0}  \\
    CDNS   & \textbf{1.4±0.24}  & 1.1±0.58  & 1.32±0.19  \\
    CMG    & 2.35±0.52 & 1.64±0.89 & \textbf{4.0±3.24}   \\
    DVN    & \textbf{0.35±3.78} & 0.18±0.1  & 0.24±0.05  \\
    DWDP   & \textbf{1.48±0.18} & 1.02±0.53 & 1.46±0.27  \\
    DXCM   & \textbf{3.3±0.32}  & 1.87±2.45 & 2.01±2.61  \\
    FANG   & \textbf{4.0±3.41}  & 0.32±0.17 & 4.0±3.52   \\
    FTNT   & \textbf{0.22±3.79} & 0.18±0.04 & 0.17±3.88  \\
    HAL    & \textbf{1.75±0.35} & 1.4±0.74  & 1.11±1.19  \\
    MKTX   & \textbf{3.36±0.48} & 1.7±0.61  & 1.33±3.07  \\
    MRO    & \textbf{4.0±0.1}   & 4.0±0.29  & 4.0±0.03   \\
    NVDA   & 0.0±0.0   & 0.0±0.0   & \textbf{0.0±4.0}   \\
    PYPL   & \textbf{0.24±0.05} & 0.13±0.05 & 0.22±3.85  \\
    SNPS   & \textbf{1.09±0.1}  & 0.89±0.35 & 0.94±3.34  \\
    SWKS   & 0.0±4.0  & 0.0±4.0  & 0.0±4.0   \\
    TSCO   & 1.72±0.12 & 1.11±0.34 & \textbf{4.0±3.03}   \\
    URI    & 0.0±0.0   & 0.0±0.0   & \textbf{4.0±4.0}   \\
    WST    & \textbf{1.79±0.36} & 1.37±0.66 & 0.98±0.33  \\
    AVB    & 0.17±0.03 & 0.09±0.03 & \textbf{4.0±3.91}   \\
    BXP    & \textbf{1.99±0.27} & 1.25±0.5  & 1.2±0.66   \\
    CINF   & \textbf{1.39±2.7}  & 1.34±3.34 & 0.89±0.57  \\
    COTY   & 0.0±0.0   & 0.0±0.0   & \textbf{0.0±4.0}   \\
    DAL    & \textbf{1.11±0.41} & 0.68±0.28 & 1.03±3.04  \\
    HWM    & \textbf{0.75±3.29} & 0.71±3.5  & 0.69±3.5   \\
    JWN    & \textbf{4.0±0.07}  & 3.58±1.06 & 4.0±0.09   \\
    LB     & \textbf{0.8±0.08}  & 0.71±0.31 & 0.78±3.29  \\
    LOW    & \textbf{4.0±0.17}  & 2.99±1.97 & 3.62±1.17  \\
    PVH    & 0.0±0.0   & 0.0±0.0   & \textbf{4.0±4.0}   \\
    RL     & 0.0±0.0   & 0.0±-0.26 & \textbf{4.0±4.0}   \\
    SLG    & \textbf{1.73±2.45} & 1.45±0.7  & 1.48±1.05  \\
    SPG    & \textbf{1.97±0.68} & 1.47±0.41 & 0.92±0.78  \\
    TPR    & \textbf{0.06±0.02} & 0.05±0.02 & 0.05±3.97  \\
    WFC    & 0.0±0.0   & 0.0±4.0  & 0.0±4.0  
      \end{tabular}
  \caption{Comparison of normalized average returns learned for three different reward functions with best performing values highlighted in bold. TradeR demonstrates higher and stable returns with balance as a reward metric.}
\end{table}

\begin{figure}[H]
  \centering
  \includegraphics[height=22cm,width=17.5cm]{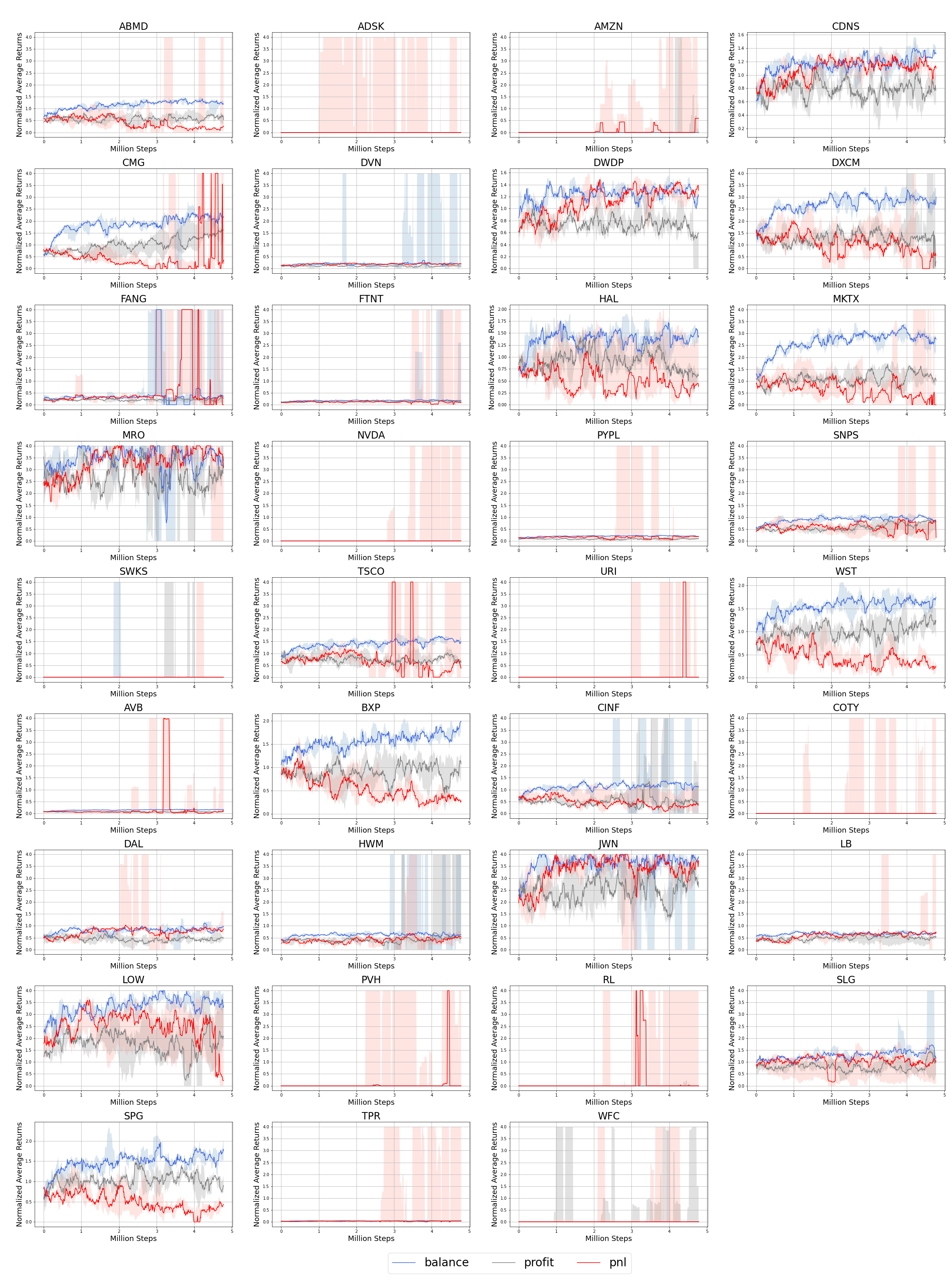}
  \caption{Normalized average returns for three different reward functions. While profits and PnL differences present high variance and in some cases divergence, balance demonstrates stability in learning serving as a suitable reward metric.}
\end{figure}

\subsection{Energy-based Intrinsic Motivation}

\begin{table}[H]
  \centering
  \begin{tabular}{l|l|l}
    Symbol & Without Energy   & With Energy        \\
    \hline
    ABMD   & 1.13±0.12    & \textbf{1.24±0.16}  \\
    ADSK   & 1.15±0.18    & \textbf{1.22±0.19}  \\
    AMZN   & \textbf{1.25±0.2}     & 1.25±0.32  \\
    CDNS   & \textbf{1.3±0.24}     & 1.22±0.22  \\
    CMG    & 1.13±0.19    & \textbf{1.2±0.09}   \\
    DVN    & 1.18±0.45    & \textbf{1.21±0.17}  \\
    DWDP   & 1.05±0.09    & \textbf{1.11±0.07}  \\
    DXCM   & \textbf{1.28±0.91}    & 1.25±0.26  \\
    FANG   & 1.17±0.3     & \textbf{1.25±0.27}  \\
    FTNT   & 1.22±0.93    & \textbf{1.24±0.23}  \\
    HAL    & \textbf{1.29±0.38}    & 1.25±0.61  \\
    MKTX   & \textbf{1.12±0.12}    & 1.11±0.06  \\
    MRO    & 1.23±0.8     & \textbf{1.25±0.28}  \\
    NVDA   & 1.19±0.23    & \textbf{1.25±0.29}  \\
    PYPL   & 1.2±0.3      & \textbf{1.3±0.35}   \\
    SNPS   & 1.13±0.13    & \textbf{1.21±0.22}  \\
    SWKS   & 1.15±0.2     & \textbf{1.25±0.18}  \\
    TSCO   & \textbf{1.29±0.64}    & 1.26±0.31  \\
    URI    & 1.17±0.17    & \textbf{1.28±0.18}  \\
    WST    & 1.18±0.25    & \textbf{1.26±0.16}  \\
    AVB    & \textbf{1.17±0.19}    & 1.16±0.13  \\
    BXP    & 1.11±0.1     & \textbf{1.18±0.18}  \\
    CINF   & 1.08±0.22    & \textbf{1.19±0.15}  \\
    COTY   & \textbf{1.21±0.21}    & 1.2±0.29   \\
    DAL    & 1.18±0.6     & \textbf{1.28±0.31}  \\
    HWM    & 1.02±0.09    & \textbf{1.08±0.04}  \\
    JWN    & \textbf{1.45±0.89}    & 1.19±0.39  \\
    LB     & 1.19±0.22    & \textbf{1.27±0.29}  \\
    LOW    & 1.18±0.26    & \textbf{1.32±0.17}  \\
    PVH    & 1.19±0.2     & \textbf{1.39±0.79}  \\
    RL     & 1.15±0.2     & \textbf{1.19±0.18}  \\
    SLG    & 1.11±0.96    & \textbf{1.23±0.41}  \\
    SPG    & 1.15±0.19    & \textbf{1.16±0.3}   \\
    TPR    & 1.2±0.25     & \textbf{1.28±0.18}  \\
    WFC    & 1.15±0.16    & \textbf{1.28±0.89} 
              \end{tabular}
  \caption{Comparison of normalized average returns with and without the presence of energy-based intrinsic motivation scheme. TradeR demonstrates improved returns as a result of surprise minimizing energy-based scheme. }
\end{table}

\begin{figure}[H]
  \centering
  \includegraphics[height=22cm,width=17.5cm]{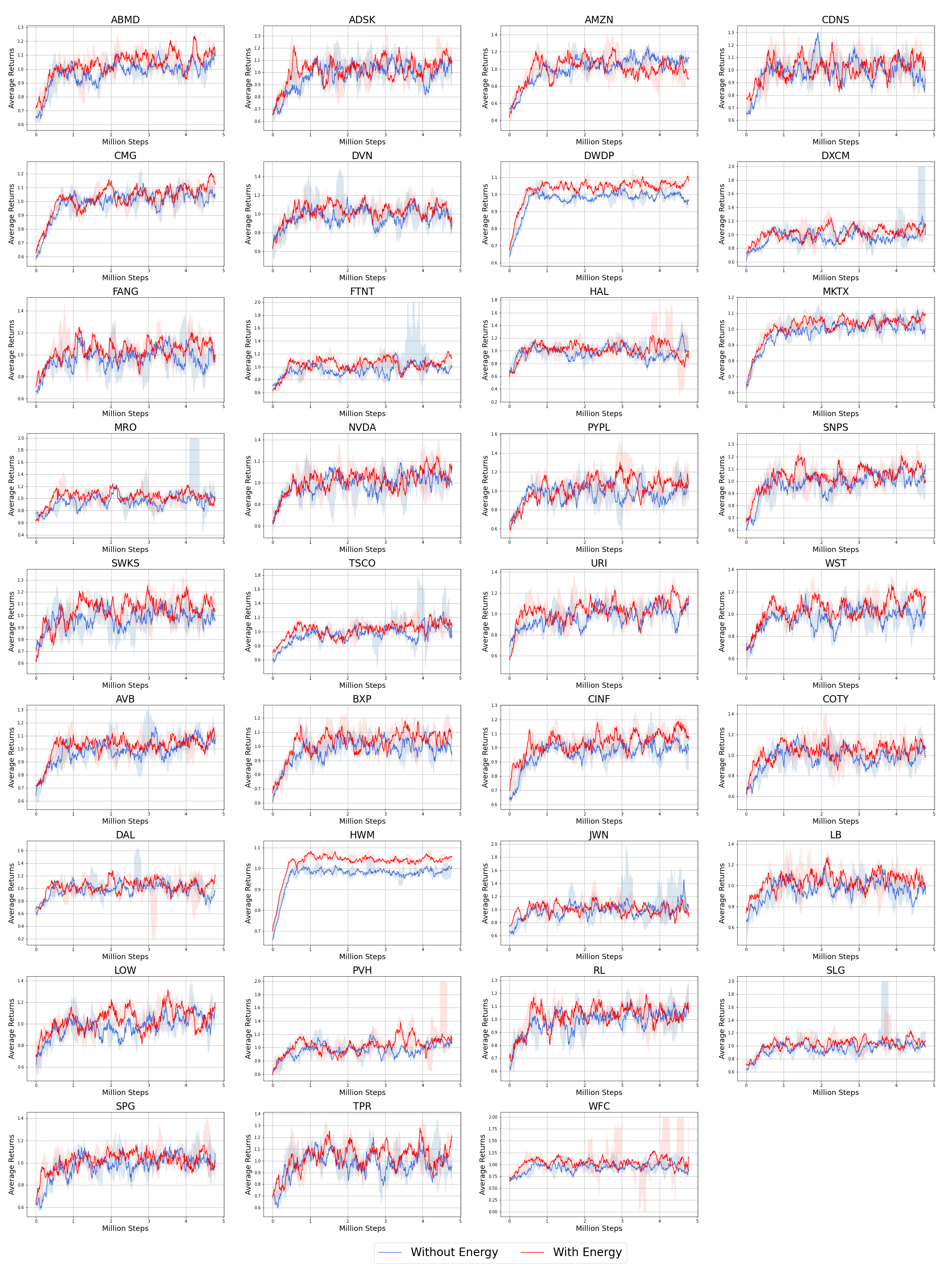}
  \caption{Normalized average returns with and without energy-based Intrinsic Motivation (IM) scheme. TradeR presents sample-efficient learning and consistent performance across its random seeds as a result of energy-based surprise minimization. }
\end{figure}

\begin{table}[H]
  \centering
  \begin{tabular}{l|l|l}
    Symbol & Without Energy   & With Energy        \\
    \hline
    ABMD   & 2.05±0.31    & \textbf{1.57±0.08}  \\
    ADSK   & 2.08±0.02    & \textbf{1.54±0.19}  \\
    AMZN   & 1.83±0.06    & \textbf{1.38±0.14}  \\
    CDNS   & 1.97±0.17    & \textbf{1.36±0.04}  \\
    CMG    & 2.28±0.29    & \textbf{1.74±0.23}  \\
    DVN    & 1.9±0.38     & \textbf{1.6±0.37}   \\
    DWDP   & 1.73±0.31    & \textbf{1.22±0.12}  \\
    DXCM   & 2.07±0.28    & \textbf{1.66±0.17}  \\
    FANG   & 2.05±0.48    & \textbf{1.61±0.17}  \\
    FTNT   & 2.04±0.18    & \textbf{1.48±0.14}  \\
    HAL    & 2.22±0.16    & \textbf{1.57±0.5}   \\
    MKTX   & 2.25±0.29    & \textbf{1.53±0.08}  \\
    MRO    & 2.51±0.37    & \textbf{1.63±0.33}  \\
    NVDA   & 2.19±0.26    & \textbf{1.75±0.21}  \\
    PYPL   & 2.05±0.21    & \textbf{1.54±0.07}  \\
    SNPS   & \textbf{1.47±0.17}    & 1.85±0.17  \\
    SWKS   & 1.92±0.26    & \textbf{1.47±0.14}  \\
    TSCO   & 2.3±0.27     & \textbf{1.81±0.11}  \\
    URI    & 1.8±0.09     & \textbf{1.27±0.11}  \\
    WST    & 2.08±0.46    & \textbf{1.46±0.09}  \\
    AVB    & 2.19±0.34    & \textbf{1.53±0.03}  \\
    BXP    & 1.96±0.28    & \textbf{1.54±0.18}  \\
    CINF   & 1.93±0.07    & \textbf{1.36±0.12}  \\
    COTY   & \textbf{1.52±0.26}    & 1.98±0.25  \\
    DAL    & 2.12±0.24    & \textbf{1.62±0.17}  \\
    HWM    & 1.93±0.13    & \textbf{1.39±0.04}  \\
    JWN    & 2.44±0.18    & \textbf{1.8±0.03}   \\
    LB     & 2.25±0.43    & \textbf{1.72±0.21}  \\
    LOW    & 1.91±0.21    & \textbf{1.32±0.09}  \\
    PVH    & 2.18±0.22    & \textbf{1.49±0.18}  \\
    RL     & 2.14±0.37    & \textbf{1.57±0.16}  \\
    SLG    & 2.37±0.19    & \textbf{1.78±0.25}  \\
    SPG    & 2.0±0.46     & \textbf{1.38±0.15}  \\
    TPR    & 2.33±0.51    & \textbf{1.41±0.44}  \\
    WFC    & 2.24±0.31    & \textbf{1.56±0.12} 
          \end{tabular}
  \caption{Comparison of normalized average shares sold with and without the energy-based scheme. TradeR suitably minimizes the amount of sold assests in order to obtain a long-term rewarding strategy. }
\end{table}

\begin{figure}[H]
  \centering
  \includegraphics[height=22cm,width=17.5cm]{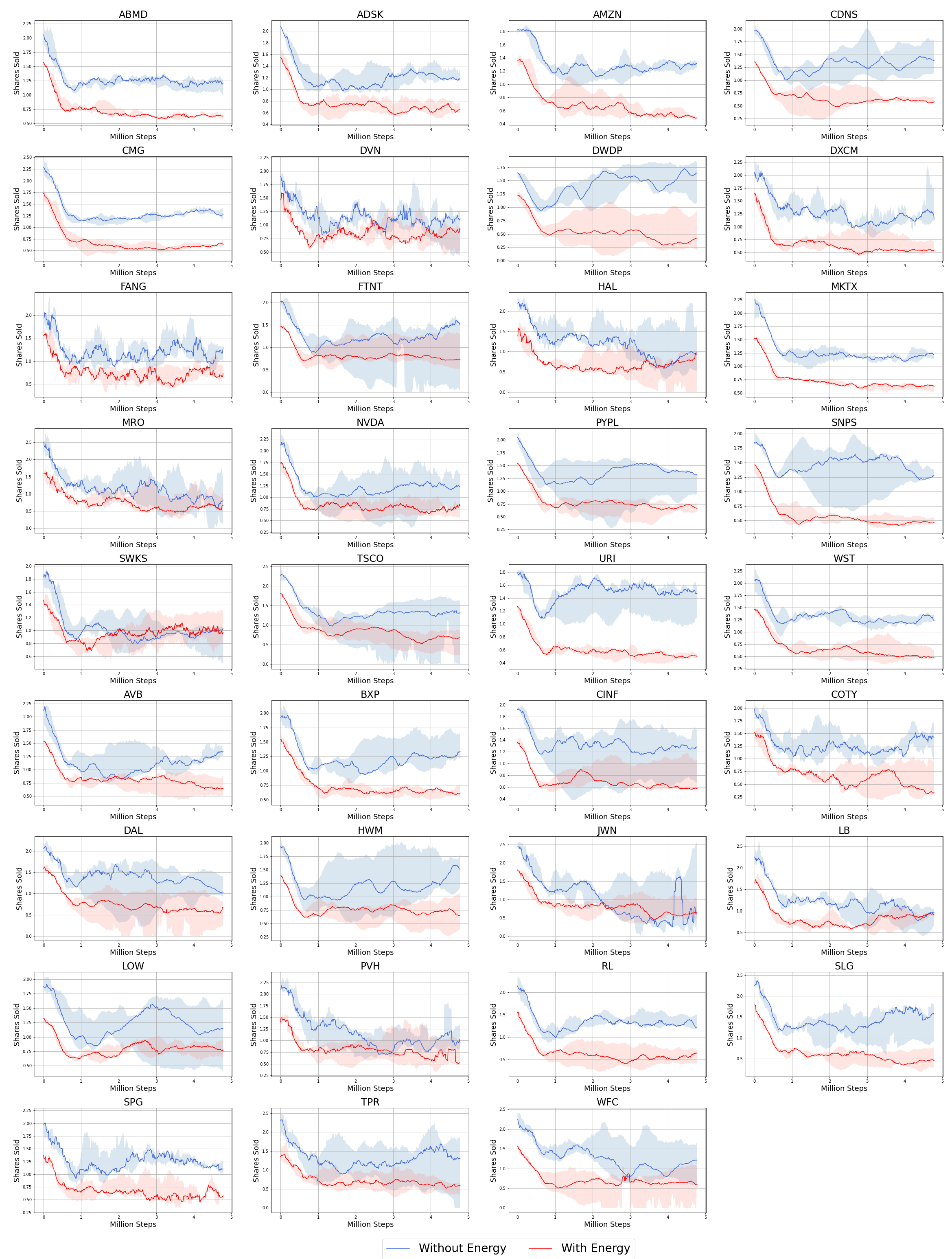}
  \caption{Normalized average shares sold by TradeR agent with and without the provision of energy-based Intrinsice Motivation (IM) scheme. The agent is able to reduce the number of sold assets by obeying the practical \textit{buy more sell less} strategy in the presence of surprise minimization. }
\end{figure}

\subsection{Hold Action}

\begin{table}[H]
  \centering
  \begin{tabular}{l|l|l}
    Symbol & Without Hold & With Hold  \\
    \hline
    ABMD   & 1.13±0.12    & \textbf{1.42±0.84}  \\
    ADSK   & 1.13±0.18    & \textbf{1.25±0.17}  \\
    AMZN   & \textbf{1.28±0.21}    & 1.23±0.2   \\
    CDNS   & 1.28±0.23    & \textbf{1.32±0.52}  \\
    CMG    & 1.13±0.19    & \textbf{1.14±0.12}  \\
    DVN    & 1.17±0.45    & \textbf{1.4±0.44}   \\
    DWDP   & 1.05±0.09    & \textbf{1.12±0.06}  \\
    DXCM   & \textbf{1.29±0.91}    & 1.22±0.22  \\
    FANG   & 1.18±0.31    & \textbf{1.36±0.34}  \\
    FTNT   & 1.23±0.92    & \textbf{1.26±0.21}  \\
    HAL    & 1.28±0.38    & \textbf{1.29±0.45}  \\
    MKTX   & 1.13±0.12    & \textbf{1.16±0.1}   \\
    MRO    & 0.97±1.05    & \textbf{1.05±1.01}  \\
    NVDA   & 1.17±0.23    & \textbf{1.24±0.21}  \\
    PYPL   & 1.2±0.3      & \textbf{1.25±0.38}  \\
    SNPS   & 1.14±0.13    & \textbf{1.22±0.08}  \\
    SWKS   & 1.15±0.2     & \textbf{1.27±0.15}  \\
    TSCO   & \textbf{1.3±0.65}     & 1.18±0.2   \\
    URI    & 1.17±0.17    & \textbf{1.24±0.19}  \\
    WST    & 1.17±0.25    & \textbf{1.27±0.23}  \\
    AVB    & \textbf{1.15±0.18}    & 1.16±0.12  \\
    BXP    & 1.11±0.1     & \textbf{1.2±0.19}   \\
    CINF   & 1.07±0.22    & \textbf{1.19±0.22}  \\
    COTY   & 1.19±0.21    & \textbf{1.3±0.2}    \\
    DAL    & 1.18±0.6     & \textbf{1.24±0.39}  \\
    HWM    & \textbf{1.01±0.09}    & 1.1±0.67   \\
    JWN    & \textbf{1.41±0.87}    & 1.32±0.84  \\
    LB     & 1.19±0.22    & \textbf{1.25±0.35}  \\
    LOW    & 1.16±0.25    & \textbf{1.25±0.2}   \\
    PVH    & 1.18±0.2     & \textbf{1.27±0.86}  \\
    RL     & 1.13±0.2     & \textbf{1.2±0.15}   \\
    SLG    & 1.12±0.95    & \textbf{1.22±0.28}  \\
    SPG    & 1.14±0.19    & \textbf{1.23±0.38}  \\
    TPR    & 1.15±0.24    & \textbf{1.27±0.84}  \\
    WFC    & 1.17±0.17    & \textbf{1.27±0.24} 
          \end{tabular}
  \caption{Comparison of normalized average returns learned with and without the hold action. Provision of holding on to current assets allows the TradeR agent to obtain rewarding payoffs in the long-horizon. }
\end{table}

\begin{figure}[H]
  \centering
  \includegraphics[height=22cm,width=17.5cm]{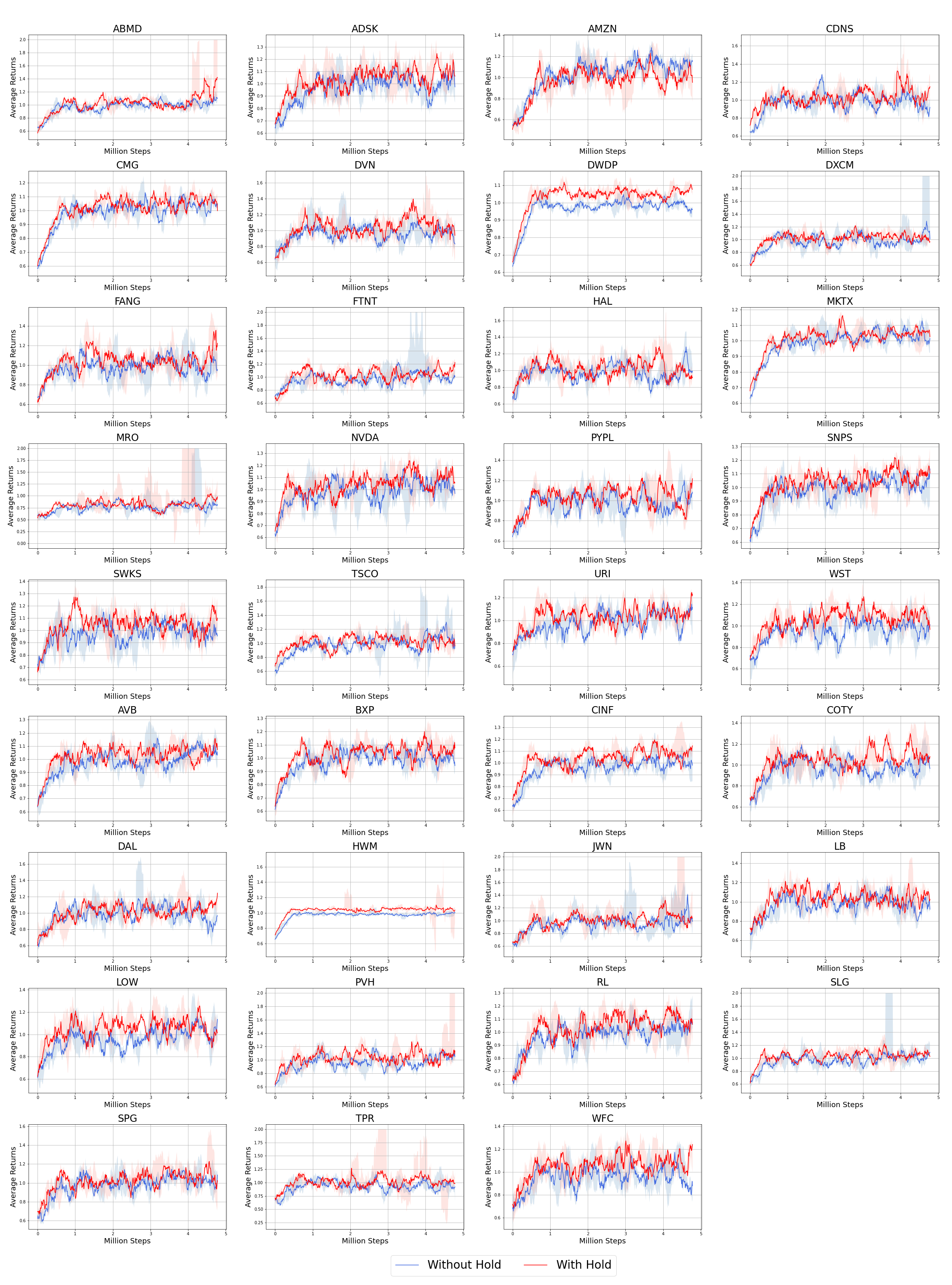}
  \caption{Normalized average returns for TradeR agent with and without the hold action. Utilization of hold action allows TradeR agent to wisely execute order bids by estimating long-term price variations. }
\end{figure}

\begin{table}[H]
  \centering
  \begin{tabular}{l|l|l}
    Symbol & Without Hold & With Hold  \\
    \hline
    ABMD   & 2.17±0.33    & \textbf{1.56±0.25}  \\
    ADSK   & 2.24±0.03    & \textbf{1.59±0.06}  \\
    AMZN   & 1.78±0.06    & \textbf{1.27±0.08}  \\
    CDNS   & 2.08±0.18    & \textbf{1.53±0.26}  \\
    CMG    & 2.27±0.29    & \textbf{1.66±0.28}  \\
    DVN    & 2.07±0.41    & \textbf{1.61±0.24}  \\
    DWDP   & 1.72±0.31    & \textbf{1.21±0.13}  \\
    DXCM   & 2.06±0.27    & \textbf{1.59±0.11}  \\
    FANG   & 2.18±0.51    & \textbf{1.46±0.28}  \\
    FTNT   & 2.25±0.2     & \textbf{1.67±0.14}  \\
    HAL    & 2.09±0.15    & \textbf{1.65±0.31}  \\
    MKTX   & 2.21±0.28    & \textbf{1.6±0.15}   \\
    MRO    & 2.61±0.38    & \textbf{1.49±1.05}  \\
    NVDA   & 2.14±0.26    & \textbf{1.61±0.21}  \\
    PYPL   & 2.11±0.21    & \textbf{1.48±0.1}   \\
    SNPS   & \textbf{1.36±0.15}    & 1.6±0.1    \\
    SWKS   & 2.22±0.3     & \textbf{1.71±0.14}  \\
    TSCO   & 2.39±0.28    & \textbf{1.87±0.17}  \\
    URI    & 1.77±0.09    & \textbf{1.2±0.14}   \\
    WST    & 2.07±0.46    & \textbf{1.48±0.06}  \\
    AVB    & 2.21±0.34    & \textbf{1.49±0.08}  \\
    BXP    & \textbf{1.4±0.25}    & 1.75±0.11   \\
    CINF   & 2.08±0.08    & \textbf{1.52±0.14}  \\
    COTY   & \textbf{1.48±0.24}     & 1.8±0.08  \\
    DAL    & 1.97±0.23    & \textbf{1.54±0.12}  \\
    HWM    & 2.18±0.14    & \textbf{1.53±0.06}  \\
    JWN    & 2.31±0.17    & \textbf{1.78±0.06}  \\
    LB     & 2.39±0.46    & \textbf{1.77±0.23}  \\
    LOW    & 2.04±0.23    & \textbf{1.41±0.19}  \\
    PVH    & 2.21±0.23    & \textbf{1.48±0.19}  \\
    RL     & 2.08±0.36    & \textbf{1.5±0.22}   \\
    SLG    & 2.14±0.17    & \textbf{1.42±0.11}  \\
    SPG    & 2.08±0.48    & \textbf{1.63±0.07}  \\
    TPR    & 2.61±0.58    & \textbf{1.61±0.1}   \\
    WFC    & 2.06±0.29    & \textbf{1.36±0.09} 
              \end{tabular}
  \caption{Comparison of normalized average returns learned by TradeR agent with and without the provision of hold action. The agent suitably minimizes the number of sold assets to obtain surprise-agnostic strategies.}
\end{table}

\begin{figure}[H]
  \centering
  \includegraphics[height=22cm,width=17.5cm]{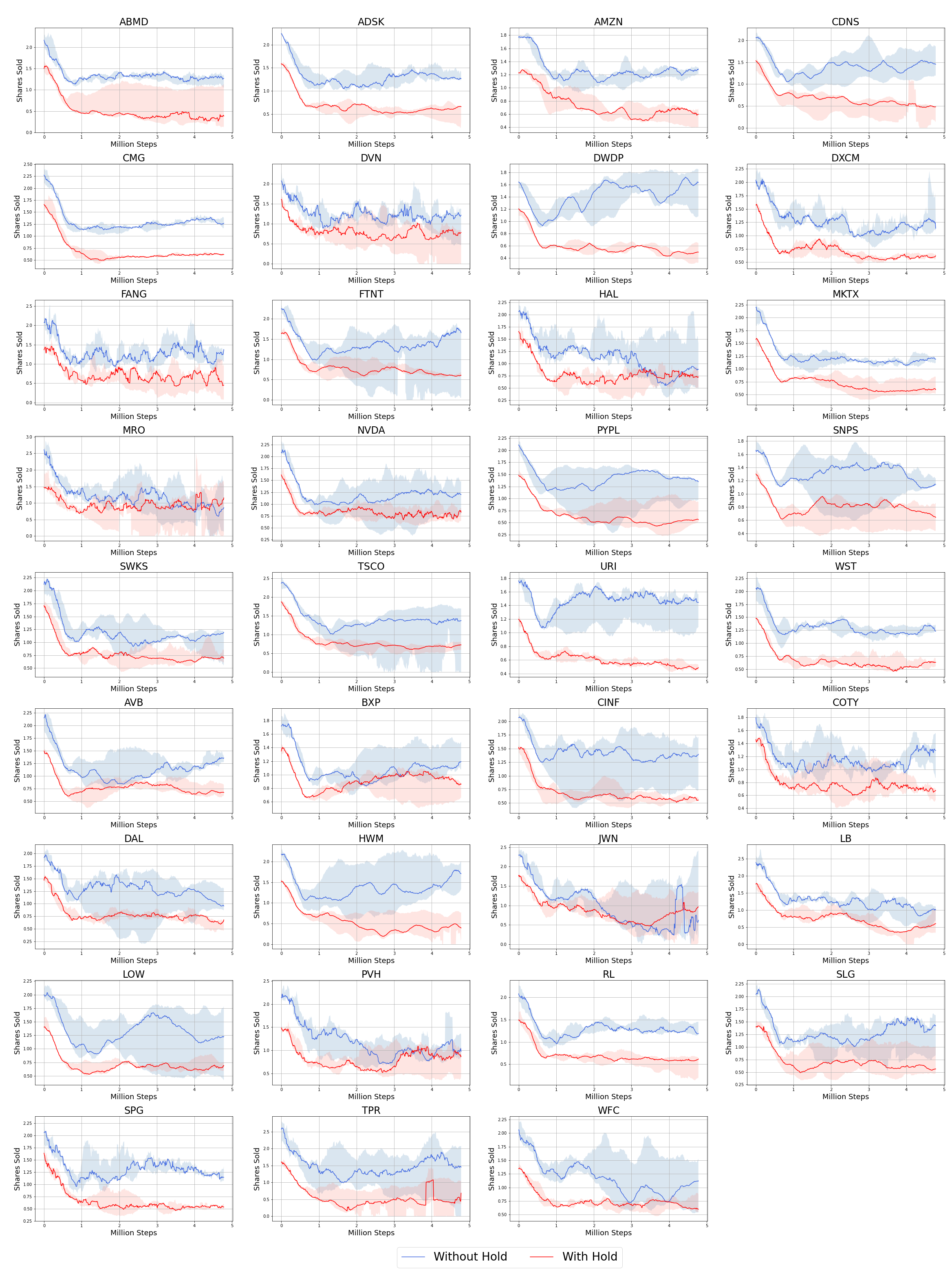}
  \caption{Normalized average shares sold by TradeR agent with and without the provision of hold action. The agent is able to reduce the number of sold assets by obeying the practical \textit{buy more sell less} strategy in the presence of hold action. }
\end{figure}

\subsection{Starting Balance}

\begin{table}[H]
  \centering
  \begin{tabular}{l|l|l|l}
    Symbol & 10,000    & 20,000    & 50,000     \\
    \hline
    ABMD   & 0.51±0.04 & 0.89±0.12 & \textbf{2.08±0.21}  \\
    ADSK   & 0.51±0.09 & 0.95±0.04 & \textbf{2.11±0.33}  \\
    AMZN   & 1.14±0.32 & 1.13±0.29 & \textbf{1.43±0.23}  \\
    CDNS   & 0.44±0.07 & 0.89±0.12 & \textbf{2.42±0.44}  \\
    CMG    & 0.82±0.1  & 0.98±0.17 & \textbf{1.63±0.27}  \\
    DVN    & 0.44±0.51 & 0.89±0.19 & \textbf{2.23±0.86}  \\
    DWDP   & 0.4±0.03  & 0.82±0.04 & \textbf{2.01±0.17}  \\
    DXCM   & 0.51±0.11 & 0.89±0.2  & \textbf{2.35±2.15}  \\
    FANG   & 0.57±1.36 & 0.91±0.3  & \textbf{2.21±0.57}  \\
    FTNT   & 0.46±0.11 & 0.87±0.12 & \textbf{2.31±2.44}  \\
    HAL    & 0.44±0.07 & 0.9±4.2   & \textbf{2.42±0.72}  \\
    MKTX   & 0.58±0.03 & 0.86±0.06 & \textbf{1.92±0.21}  \\
    MRO    & 0.48±0.32 & 0.99±0.39 & \textbf{2.33±2.73}  \\
    NVDA   & 0.57±0.1  & 0.91±0.19 & \textbf{2.16±0.43}  \\
    PYPL   & 0.52±0.1  & 0.89±0.29 & \textbf{2.26±0.56}  \\
    SNPS   & 0.52±0.09 & 0.91±0.14 & \textbf{2.09±0.25}  \\
    SWKS   & 0.48±0.11 & 0.87±0.18 & \textbf{2.19±0.38}  \\
    TSCO   & 0.51±0.07 & 0.87±0.12 & \textbf{2.42±1.21}  \\
    URI    & 0.49±0.15 & 0.88±0.13 & \textbf{2.2±0.32}   \\
    WST    & 0.48±0.08 & 0.96±0.1  & \textbf{2.21±0.47}  \\
    AVB    & 0.52±0.06 & 0.88±0.11 & \textbf{2.11±0.34}  \\
    BXP    & 0.47±0.08 & 0.87±0.45 & \textbf{2.08±0.2}   \\
    CINF   & 0.46±0.88 & 0.87±1.25 & \textbf{2.02±0.41}  \\
    COTY   & 0.43±0.07 & 0.91±0.3  & \textbf{2.29±0.41}  \\
    DAL    & 0.48±0.09 & 0.9±0.18  & \textbf{2.21±1.13}  \\
    HWM    & 0.38±4.63 & 0.76±0.03 & \textbf{1.87±0.16}  \\
    JWN    & 0.45±0.14 & 0.87±0.22 & \textbf{2.76±1.7}   \\
    LB     & 0.45±0.1  & 0.95±0.22 & \textbf{2.29±0.42}  \\
    LOW    & 0.49±0.1  & 0.91±0.16 & \textbf{2.2±0.48}   \\
    PVH    & 0.47±0.06 & 1.01±0.35 & \textbf{2.25±0.38}  \\
    RL     & 0.47±0.07 & 0.89±0.16 & \textbf{2.16±0.38}  \\
    SLG    & 0.45±0.05 & 0.88±0.33 & \textbf{2.12±2.68}  \\
    SPG    & 0.47±0.12 & 0.94±0.22 & \textbf{2.12±0.36}  \\
    TPR    & 0.45±0.07 & 0.91±0.12 & \textbf{2.28±0.48}  \\
    WFC    & 0.43±0.09 & 1.81±2.67 & \textbf{2.2±0.31}  
        \end{tabular}
  \caption{Comparison of normalized average returns learned with increasing amount of seed balance provided to the TradeR agent. The agent suitably demonstrates optimal performance for higher starting balance values indicating an apt utilization of assets. }
\end{table}

\begin{figure}[H]
  \centering
  \includegraphics[height=22cm,width=17.5cm]{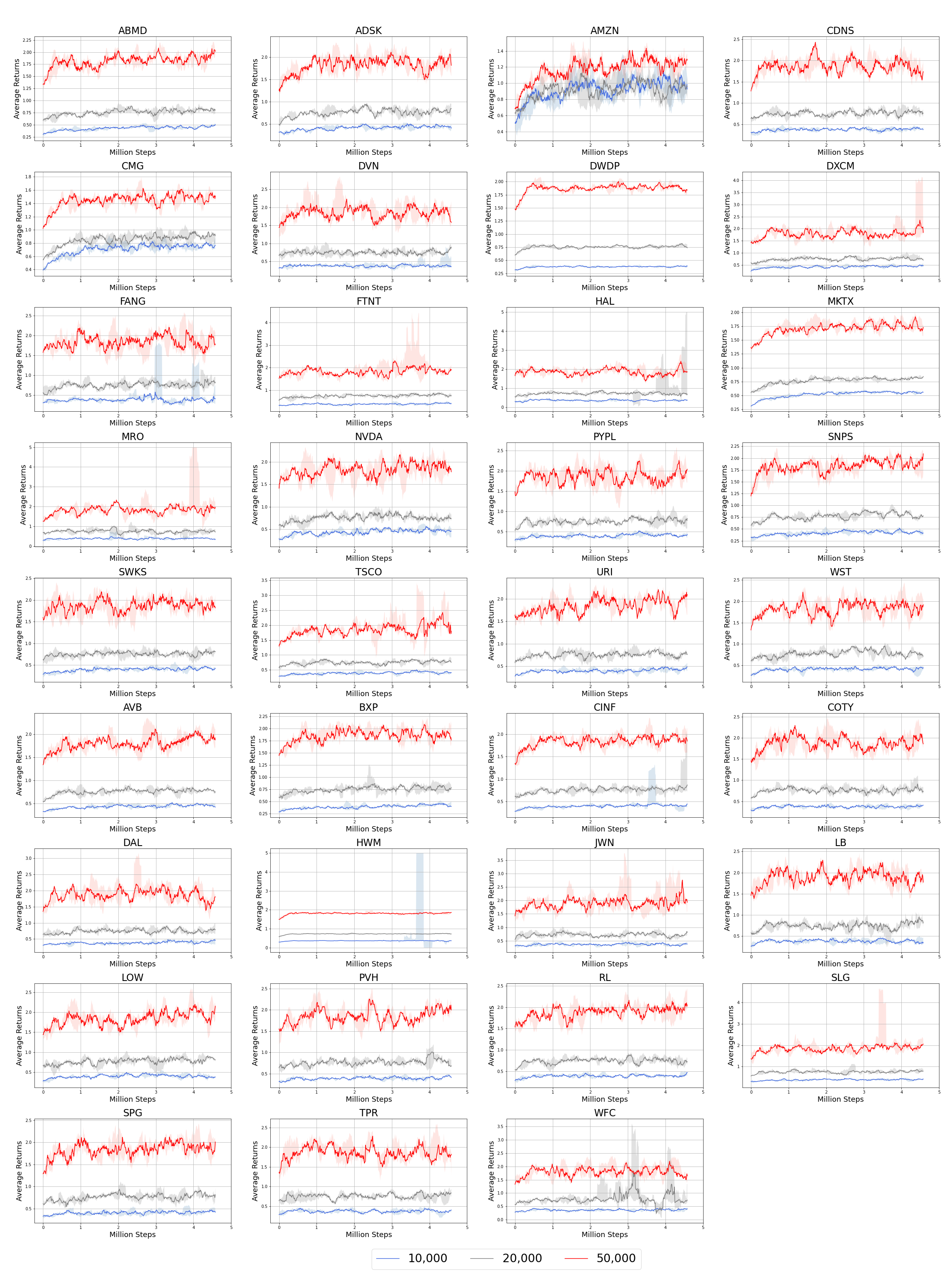}
  \caption{Normalized average returns for three different starting balance values. TradeR suitably leverages large seed values which allow the agent to attain higher payoffs.}
\end{figure}

\end{document}